\documentclass[journal]{IEEEtran}
\usepackage{amsthm}
\usepackage{bbm}
\usepackage{amssymb}
\usepackage{graphicx}
\usepackage{color}
\usepackage{float}
\usepackage{cite}
\usepackage{algorithm}
\usepackage{algorithmic}
\usepackage{amsmath}
\newtheorem{theorem}{Theorem}
\newtheorem{lemma}{Lemma}
\newtheorem{definition}{Definition}

\newtheorem{corollary}{Corollary}

\allowdisplaybreaks
\begin{document}
	\newcommand{\tabincell}[2]{\begin{tabular}{@{}#1@{}}#2\end{tabular}}
	\title{Scheduling to Minimize Age of Synchronization in Wireless Broadcast Networks with Random Updates}
	
\author{Haoyue~Tang,~\IEEEmembership{Student Member,~IEEE},~Jintao Wang,~\IEEEmembership{Senior Member,~IEEE},~Zihan~Tang,~Jian~Song,~\IEEEmembership{Fellow,~IEEE}
		\thanks{Manuscript received March 21, 2019, revised September 1, 2019 and November 24, 2019, accepted Feburary 20, 2020. The authors are with the Department of Electronic Engineering, Tsinghua University, Beijing 100084, China and Beijing National Research Center for Information Science and Technology (BNRist). J. Wang and J. Song are also with Research Institute of Tsinghua University in Shenzhen, Shenzhen, 518057. (e-mail: tanghaoyue13@tsinghua.org.cn; wangjintao@tsinghua.edu.cn; tangzh14@mails.tsinghua.edu.cn; jsong@tsinghua.edu.cn). This work has been presented in part at the 2019 IEEE International Symposium on Information Theory \cite{haoyue}. \emph{(Corresponding author: Jintao Wang)}}
		\thanks{This work was supported in part by the National Key R\&D Program of China under Grant 2017YFE0112300, Beijing National Research Center for Information Science and Technology under Grant BNR2019RC01014 and BNR2019TD01001 and the Tsinghua University Tutor Research Fund. }}
	
	\maketitle
	\begin{abstract}
	In this work, a wireless broadcast network with a base station (BS) sending random time-sensitive information updates to multiple users under bandwidth constraint is considered. To measure data desynchronization between BS and the user, the metric Age of Synchronization (AoS) is adopted in this work. It shows the amount of the time elapsed since freshest information at the receiver becomes desynchronized.  The AoS minimization scheduling problem is formulated into a discrete time Markov decision process and the optimal solution is approximated through structural finite state policy iteration. An index based heuristic scheduling policy based on restless multi-arm bandit (RMAB) is provided to further reduce the computational complexity. Simulation results show that the proposed index policy achieves compatible performance with the MDP and is close to the AoS lower bound. Our work indicates that, to obtain a small AoS over the entire network, users with larger transmission success probability and smaller random update probability are more likely to be scheduled at smaller AoS.
		
		
	\end{abstract}

	\section{Introduction}
	
	The design of next generation mobile and wireless communication networks are driven partly by the need of mission-critical services like real-time control and the Internet of Things (IoT). Moreover, the proliferation of mobile devices have boosted the need to enhance the timeliness of services like instant chatting, mobile ads, social updates notifications, etc. The above applications require that each user possesses fresh data about the information they are interested in.
	
	To measure data freshness from the perspective of the receiver when update packets are generated randomly by external actions or environments, e.g., in databases and error alarm systems, the metric called \emph{Age of Synchronization} (AoS) is proposed \cite{zhong_two_2018}. By definition, AoS measures the time elapsed since freshest information at the receiver becomes \emph{desynchronized}. Compared with the metric called Age of Information (AoI) which measures the time elapsed since the freshest information at the receiver is \textit{generated}, the AoS accounts for whether the source being tracked has actually changed, while AoI measures the combination of the content update inter-generation duration and content desynchronization. To better explain their differences, consider a database synchronization problem as an example. Suppose the file stored at the receiver is the same as the one in remote databases, which has not been changed or updated by external users or environment for a long time. By definition, AoS equals 0 because files at the receiver and at the remote database are synchronized. However, AoI can be very large due to the long interval between two remote database updates and thus file desynchronization status cannot be inferred directly from AoI. On the contrary, AoS is a more appropriate metric in studying database desynchronization. Similar scenarios can be found in monitoring system \cite{erroralarm} and web crawling problems \cite{webcrawling}. Due to the aforementioned differences between AoI and AoS, scheduling strategies that aim at minimizing AoI may not guarantee a good AoS performance. Thus, it is of importance to study scheduling strategies to obtain a good AoS performance.
	
	Data freshness optimization has received lots of attention in communication system design. The problem of minimizing AoI have been investigated in coding \cite{mayekar_isit_2018,parag_wcnc_2017,zhong_dcc_2016}, physical layer design \cite{klein_twc_2017,costa_icc_2015} and network optimization \cite{igor16allerton,igor_ton_2018,yates_cache,igor18infocom,talak2017allerton,talak_distributed_18,hsu17isit,hsu_18_isit,jiang_isit_2018,lu_age_2018}. When the source keeps changing all the time and the update packets carrying those updates can be generated at will, centralized scheduling algorithm to optimize AoI performance in networks with interference constraint is first studied in \cite{igor16allerton}. Theoretic lower bound for AoI performance is derived and various scheduling policies are proposed to approach the bound \cite{igor_ton_2018}. When the generation of update packets cannot be controlled at will and appear in a stochastic manner, theoretic performance and scheduling algorithms have been studied in \cite{hsu17isit,hsu_18_isit,lu_age_2018,jiang_isit_2018,sun2019optimizing}. However, scheduling policies provided in \cite{hsu17isit,hsu_18_isit,lu_age_2018,jiang_isit_2018} are based on error free transmission, while in practical wireless communication scenarios, packet loss may happen due to channel fading and decoding error. Moreover, those researches all focus on minimizing AoI. The metric AoS, which measures the content desynchronization alone, is a more appropriate metric to measure the desynchronization status of caches, databases and error alarm systems \cite{aos_ton,synchornization}. Although AoS has been used to measure the staleness of replicated databases and the its performance has been studied under several updating strategies in \cite{aos_ton,synchornization}, the effect of packet loss has not been taken into account. Besides, update strategies in the above works are unaware of the database changes, i.e., scheduling decision may take place even if there is no change in the database content.

	To fill this gap, we aim at designing scheduling policies to minimize the expected AoS of an unreliable wireless broadcast network, when the generation of update packets cannot be controlled at will and arrives stochastically because of external environment. Unlike previous work that consider update randomness at the transmitter but no transmission randomness at the receiver\cite{hsu17isit,hsu_18_isit,jiang_isit_2018}, we consider double randomness at both the transmitter and the receiver. Our contributions are summarized as follows:
	\begin{itemize}
		\item We derive the theoretic lower bound of the AoS performance in error-prone wireless networks when the update of each source appears following i.i.d Bernoulli distribution. 
		
		\item The AoS scheduling problem is reformulated into a Markov decision process (MDP). We exploit the switching structure of the optimum policy by analyzing the monotonic characteristic and submodularity of the value function. The optimum solution is approximated through finite state policy iteration.
		
		\item To overcome the computational complexity of the MDP solution, we propose a heuristic index based algorithm by reformulating the scheduling problem into restless multi-arm bandit (RMAB). We prove that each bandit is indexable and derive the closed form expression of the Whittle's index. Simulation results show that the Whittle's index policy can achieve AoS performance close to the MDP solution and the AoS lower bound.
	\end{itemize}

	The remainder of this paper is organized as follows. The network model and the two metrics, AoI and AoS are introduced and compared in Section~II, where the overall scheduling problem is formulated and AoS lower bound is derived. In Section~\ref{sec:MDP}, we reformulate the problem into a Markov decision process and propose a structural policy iteration to approximate the MDP solution. In Section~IV, we propose an index based algorithm based on restless multi-arm bandit. Simulations are provided in Section~V and Section~VI draws the conclusion.
	
		\emph{Notations: }Vectors are written in boldface
	letters. The probability of event $\mathcal{A}$ conditioned on $\mathcal{B}$ is denoted as Pr$(\mathcal{A}|\mathcal{B})$, the expectation with regard to random variable $X$ conditioned on random variable $Y$ is denoted as $\mathbb{E}_X[f(X)|Y]$. Vector $\mathbf{e}_i$ denotes vector with the $i$-th element being $1$ and the remaining elements take $0$.

	\section{System Overview}
	
	In this section, we introduce the system model, formulate the overall scheduling problem and derive the lower bound of the scheduling problem.
\subsection{Network Model}
 We consider a wireless broadcast network with a base station (BS) holding update information of $N$ randomly changing sources and broadcasting them to $N$ users. Consider a discrete time scenario and we use $t\in\{1, \cdots, T\}$ to denote the index of the current slot, the update of source $n$ appears independently and identically with probability  $\lambda_n\in(0, 1]$ in each time slot.\footnote{When $\lambda_n=1$, new update packets of source $n$ appear in every slot and thus the source keeps changing. AoS and AoI are the same metric in this scenario. In following analysis, we will compare our derivations with results of AoI in this special case to show the relationship of the two metrics.} Let the indicator function $\Lambda_n(t)\in\{0,1\}$ denote whether an update of source $n$ happens during slot $t$. If $\Lambda_n(t)=1$, then an update occurs during slot $t$, and then it can be broadcasted by the BS at the beginning of slot $(t+1)$. We assume each user is interested in the information from the corresponding source, i.e., user $n$ is only interested information about $n$, $n\in\{1,\cdots,N\}$. 
%

At the beginning of each slot, the BS schedules to send information updates over error-prone wireless links. Since each user is only interested in the freshest information about the corresponding source, we assume the BS only keeps the latest update of each source, i.e., new update packets will replace the older update packets for a typical source. Here we use the indicator function $u_n(t)\in\{0, 1\}$ to denote scheduling actions. If user $n$ is not scheduled then $u_n(t)=0$. If $u_n(t)=1$, the newest update from source $n$ is transmitted and user $n$ will successfully receive the packet by the end of slot $t$ if the transmission succeeds. Assume that the packet erasure is a memoryless Bernoulli process and user $n$ has a fixed channel characterized by the Bernoulli packet success probability $p_n$. An error-free acknowledgment sent from user $n$ will reach the BS instantaneously if the transmission succeeds. In each slot, the amount of data to be broadcasted must be smaller than the channel capacity and the available bandwidth (otherwise the transmission will all fail). Similar to \cite{igor_ton_2018,jiang_isit_2018,yates_cache}, in this work, we assume the BS attempts to broadcast one update in each time slot:
\begin{equation}
\sum_{n=1}^Nu_n(t)\leq 1.
\end{equation}
\subsection{Age of Information and Age of Synchronization}
To introduce the concept of AoI and AoS, we consider a single source discrete time scenario as an example. First we review the definition of them and then we talk about how AoS evolves depending on scheduling decision $\{u_n(t)\}$ and update randomness $\{\Lambda_n(t)\}$. Suppose the $i^{\text{th}}$ update packet of the source is generated during slot $g_i$. Let $r_i$ be the receiving time-stamp of the corresponding packet. If the packet is not received by the user, then denote $r_i=+\infty$.

The AoI measures the time elapsed since the latest update at the receiver is generated \cite{yates12infocom}. Let $q(t)=\max_{i\in\mathbb{N}^+}\{i|r_i< t\}$ be the index of the latest update at the beginning of slot $t$ at the receiver, which is generated in $g_{q(t)}$. By definition, the AoI $h(t)$ at the beginning of slot $t$ can be computed by:
\begin{equation}
h(t)=t-g_{q(t)}.
\end{equation}

The AoI evolution for stochastic arrivals transmitted through an unreliable communication channel can be found in \cite{sun2019optimizing}. 

\begin{figure}[h]
	\centering
	\includegraphics[width=.5\textwidth]{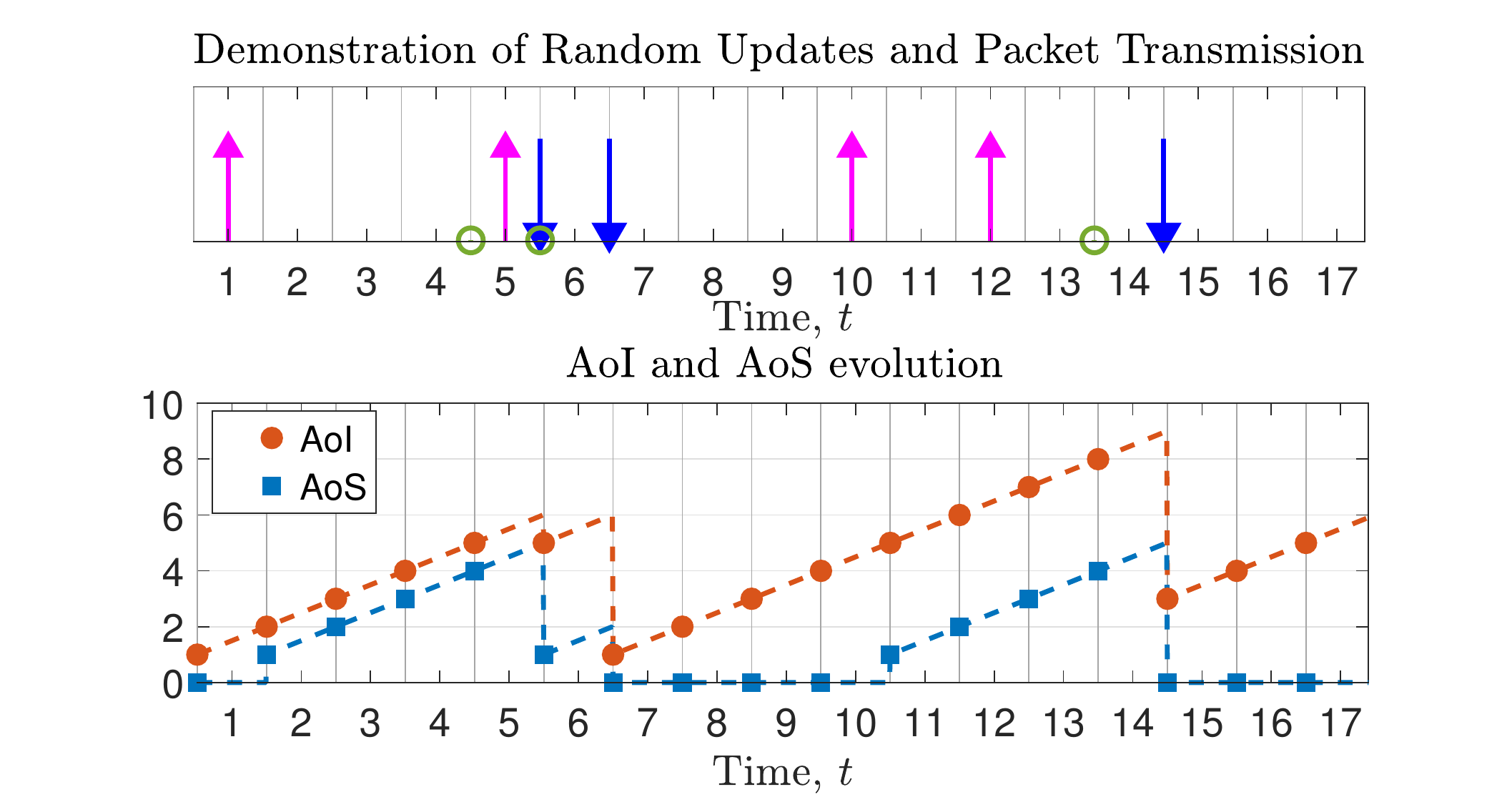}
	\caption{On the top, sample sequences representing time-stamps of update arrivals (upward magenta arrows), update sending decisions (green circles) and update received time-stamps (downward brown arrows). On the bottom, sample paths of AoI (blue) and AoS (red).}
	\label{AoSmodel}
\end{figure}

The AoS describes how long the information at the receiver has become desynchronized compared with the source \cite{zhong_two_2018}. Notice that $q(t)+1$ is the index of earliest source update since the generation time-stamp of the freshest information stored at the receiver at the beginning of slot $t$. Let $s(t)$ be the AoS at the beginning of slot $t$, by definition:
\begin{equation}
s(t)=(t-g_{q(t)+1})^+,\label{eq:AoSdef}
\end{equation}
where function $(\cdot)^+=\max\{0, \cdot\}$. According Eq.~\eqref{eq:AoSdef}, if no new update arrives after the generation time-stamp of the latest refresh of the user, i.e., $g_{q(t)+1}\geq t$, then $s(t)=0$. The sample paths of AoI and AoS of a source are depicted in Fig.~\ref{AoSmodel}. From the figure, the AoS remains zero until a new fresh update arrives and increases linearly if the content stored at the receiver becomes desynchronized with the source, while AoI keeps increasing as long as no update has been received. The difference between AoI and AoS is the reference object. The AoS measures data freshness compared to the content of the random update source and accounts for the whether the process being tracked has actually changed, while AoI measures the time difference between now and the generation time-stamp of receiver's current content.

Now we return to the multiple-user scenario and introduce the evolution of AoS. Let $s_n(t)$ be the AoS of user $n$ at the beginning of each slot $t$. The analysis is divided into two categories based on desynchronization status: 
\begin{itemize}
\item First let us consider that the information at user $n$ is synchronized with source $n$ at the beginning of slot $t$, i.e., $s_n(t)=0$, then $s_n(t+1)$ depends on whether an update occurs during slot $t$. 
\begin{itemize}
	\item[*] When there is no update,  i.e., $\Lambda_n(t)=0$, then $s_n(t+1)=0$, indicating that user $n$ is still synchronized with source $n$ at the beginning of next slot.
	\item[*] When $\Lambda_n(t)=1$, suggesting an update of source $n$ occurs in slot $t$, then information of user $n$ will become desynchronized at the beginning of next slot, i.e., $s_n(t+1)=1$.
\end{itemize}

\item If $s_n(t)\neq 0$, then user $n$ is desynchronized with source $n$ at the beginning of slot $t$. 
\begin{itemize}
\item[+]If $u_n(t)=1$ and the transmission succeeds, then the latest information of source $n$ by the end of slot $t-1$ will be received by the end of slot $t$. Then in this case:
\begin{itemize} 
\item[*]If $\Lambda_n(t)=0$, there is no update during slot $t$, thus the received information will be synchronized with the source at the beginning of next slot, i.e., $s_n(t+1)=0$.
\item[*]If $\Lambda_n(t)=1$, the received information will be out-of-date immediately at the beginning of next slot, then $s_n(t+1)=1$. 
\end{itemize}
\item[+]If the update is not transmitted $u_n(t)=0$ or the transmission fails, the update packet will not be received by user $n$, then AoS increases linearly, i.e., $s_n(t+1)=s_n(t)+1$. 
\end{itemize}
\end{itemize}
Based on the above analysis, the dynamics of AoS for user $n$ is:
	\begin{equation}
s_n(t+1)=\begin{cases}
0,&s_n(t)=0, \Lambda_n(t)=0;\\
1,&s_n(t)=0, \Lambda_n(t)=1;\\
0,&\Lambda_n(t)=0, u_n(t)=1, \text{succeeds};\\
1,&\Lambda_n(t)=1, u_n(t)=1, \text{succeeds};\\
s_n(t)+1,&\text{otherwise.}
\end{cases}
\label{stateevolution}
\end{equation}
\subsection{Problem Formulation}
The expected average AoS of all users following policy $\pi$ over a consecutive of $T$ slots can be computed as follows: \[J_T(\pi)=\mathbb{E}_\pi\left[\frac{1}{NT}\sum_{t=1}^T\sum_{n=1}^Ns_n(t)|\mathbf{s}(0)\right],\]where the vector $\mathbf{s}(t)=[s_1(t), s_2(t), \cdots, s_N(t)]^T\in\mathbb{N}^N$ denotes the AoS of all users at the beginning of slot $t$. In this work, we assume that all the sources have been synchronized initially, i.e., $\mathbf{s}(0)=0$ and omit $\mathbf{s}(0)$. 

Let $\Pi_{NA}$ denote the class of non-anticipated policies, i.e., scheduling decisions $\{u_n(t)\}$ in slot $t$ are made based on channel statistics $\{p_n\}$, the past and current AoS of all users $\{s_n(\tau)\}_{\tau<t}$. No information about the future can be used. We aim at designing a non-anticipated scheduling strategy $\pi\in\Pi_{NA}$ such that the above time-average expected AoS is minimized when $T\rightarrow\infty$. The problem considered in this paper is organized as follows:
\begin{subequations}
	\begin{equation}
	\pi^*=\arg\min_{\pi\in\Pi_{NA}}\lim_{T\rightarrow\infty}J_T(\pi), 
	\end{equation}
	\begin{equation}
		\text{ where }J_T(\pi)=\mathbb{E}_\pi\left[\frac{1}{NT}\sum_{t=1}^T\sum_{n=1}^Ns_n(t)\right],
		\label{objprimal}
	\end{equation}
	\begin{equation}
	\text{s.t. }\mathbb{E}_\pi\left[\sum_{n=1}^N u_n(t)\right]\leq 1, \forall t.\label{eq:primalconstraint}
	\end{equation}
\end{subequations}

\subsection{Lower Bound of AoS}
In this part, a lower bound to the expected average AoS performance to the above optimization problem is derived. Sample path argument is used here to characterize the AoS evolution of each user. Then, we establish the expected average AoS over the entire network when the number of time slots $T\rightarrow\infty$. By using Fatou's lemma, the lower bound is then established.

\begin{theorem}
	For a given network setup, the average AoS over the entire network is lower bounded by:
	\begin{gather}
	\begin{align}
		&\text{\rm AoS}_{\text{\rm LB}}=\\
		&\frac{1}{N}\sum_{n=1}^N\gamma_n^*\left[\frac{1}{2}\left(\frac{1}{\gamma_n^*}-\frac{1-\lambda_n}{\lambda_n}\right)^2+\frac{1}{2}\left(\frac{1}{\gamma_n^*}-\frac{1-\lambda_n}{\lambda_n}\right)\right],\nonumber
	\end{align}
	\end{gather}
	where $\gamma_n^*=\max\{1/\sqrt{\left(\frac{1-\lambda_n}{\lambda_n}\right)^2-\left(\frac{1-\lambda_n}{\lambda_n}\right)+\frac{2\mu^*N}{p_n}}, \lambda_n\}$, and $\mu^*$ is the coefficient that keeps $\sum_{n=1}^N\frac{\gamma_n^*}{p_n}=1$.
\end{theorem}
\hspace{0.2cm}\textit{Proof Sketch: }The lower bound is obtained by solving AoS minimization problem under a relaxed bandwidth constraint, i.e., multiple users can be scheduled at the same time but the time average users scheduled in each slot is still smaller than 1. The hard bandwidth constraint in every slot indicates that the derived AoS lower bound can be loose. The bound is used to evaluate the performance of our proposed algorithms. The details of the derivations will be provided in Appendix \ref{sec:thm1proof}.
\section{Markov Decision Process}\label{sec:MDP}
In this section, we design a scheduling strategy based on Markov decision process (MDP) techniques. The optimization problem Eq.~(5a-5c) can be formulated into an MDP problem with elements explained as follows:

\begin{itemize}
	\item \textbf{State space: }The state at time slot $t$ is defined to be the AoS of all the users over the entire network $\mathbf{s}(t)$. The state space is countable but infinite because of possible transmission failures.
\item \textbf{Action space: }We define the action $a(t)$ at time $t$ to be the index of the selected user. The corresponding scheduling decision can be computed by $u_n(t)=\mathbbm{1}_{(n=a(t))}, \forall n$, where $\mathbbm{1}_{(\cdot)}$ is the indicator function. Denote $a(t)=0$ if the BS chooses to be idle. The action space $\{0, 1, 2, \cdots, N\}$ is hence countable and finite.
\item \textbf{Transition probability: }Let $\text{Pr}(\mathbf{s}'|\mathbf{s}, a)$ be the transition probability from state $\mathbf{s}(t)=\mathbf{s}=[s_1, s_2, \cdots, s_N]^T$ to state $\mathbf{s}(t+1)=\mathbf{s}'=[s_1', s_2', \cdots, s_N']^T$ at the next slot by taking action $a$ at slot $t$. Since the probability of new update packet arrival and channel states are independent among the users, the transition probability can be decomposed into:
\begin{equation}\text{Pr}(\mathbf{s}'|\mathbf{s}, a)=\prod_{n=1}^N\text{Pr}(s_n
'|s_n, a),
\label{MDPprobabilitytransfer}
\end{equation}
where $\text{Pr}(s_n'|s_n, a)$ denotes the one-step transition probability of user $n$ given action $a$ and has the following expression according to Eq.~\eqref{stateevolution}:
\begin{align}&\text{Pr}(s_n'|s_n, a)=\nonumber\\
&\hspace{1cm}\begin{cases}
1,&s_n'=s_n+1, s_n\neq 0, a\neq n;\\
1-p_n,&s_n'=s_n+1, s_n\neq 0, a=n;\\
\lambda_n p_n,&s_n'=1, s_n\neq 0, a=n;\\
(1-\lambda_n)p_n, &s_n'=0, s_n\neq 0, a=n;\\
\lambda_n, &s_n'=1, s_n=0, \forall a;\\
1-\lambda_n, &s_n'=0, s_n=0, \forall a;\\
0, &\text{otherwise}.
\end{cases}\nonumber
\end{align}
\item \textbf{One-step cost: }Let $C(\mathbf{s}(t), a(t))$ be the one-step cost at state $\mathbf{s}(t)$ given action $a(t)$, which is the average AoS growth of the entire network at time $t$:
\[C(\mathbf{s}(t), a(t))=\frac{1}{N}\sum_{n=1}^Ns_n(t).\]
\end{itemize}	

The solution $\pi^*$ that minimizes the average AoS in Eq.~\eqref{objprimal} can be found by solving the MDP. Denote $J_\alpha(\mathbf{s}, \pi)$ be the $\alpha$-discounted cost following policy $\pi$ starting from state $\mathbf{s}$, i.e.,  
	\begin{equation}
\text{ where }J_\alpha(\mathbf{s},\pi)=\lim_{T\rightarrow\infty}\mathbb{E}_\pi\left[\sum_{t=1}^T\alpha^{t-1}C(\mathbf{s}(t), a(t))\right],
\end{equation}
In this section, we approximate $\pi^*$ by solving the $\alpha$-discounted cost problem when $\alpha\rightarrow1$. Define $\pi_\alpha^*$ be the optimum policy that minimizes the $\alpha$-discounted cost starting from any state $\mathbf{s}$ ans satisfies the bandwidth constraint, i.e., 
\begin{subequations}
	\begin{equation}
	\pi_\alpha^*=\arg\min_{\pi\in\Pi_{NA}}J_\alpha(\mathbf{s}, \pi),\forall \mathbf{s},
	\end{equation}
	\begin{equation}
	\text{s.t. }\mathbb{E}_\pi\left[\sum_{n=1}^N u_n(t)\right]\leq 1, \forall t.\label{discountprimalconstraint}
	\end{equation}
\end{subequations}

Policy $\pi_\alpha^*$ is obtained a modified policy iteration that utilizes its structure. To analyze the structure of $\pi_\alpha^*$, let us first provide the formal definition of the stationary deterministic policies:
\begin{definition}
Let $\Pi_{\text{SD}}$ denote the class of stationary deterministic policies. Given state $\mathbf{s}(t)=\mathbf{s}$, a stationary deterministic policy $\pi_{\text{SD}}\in\Pi_{\text{SD}}$ selects action $a(t)=f(\mathbf{s})$, where function $f(\cdot):\mathbf{s}\rightarrow\{0,\cdots, N\}$ is a deterministic mapping from state space to action space. 
\end{definition}

According to \cite{Bertsekas_DP}, $\pi_\alpha^*$ can be a stationary deterministic policy and denote $\pi_\alpha(\mathbf{s})$ be the action it takes in state $\mathbf{s}$. Then denote $V_\alpha(\mathbf{s})$ is the $\alpha$-discounted cost following policy $\pi_\alpha^*$ starting from state $\mathbf{s}$, i.e., 
\begin{equation}
V_\alpha(\mathbf{s})=\lim_{T\rightarrow\infty}\mathbb{E}_{\pi_\alpha^*}\left[\sum_{t=1}^T\alpha^{t-1}C(\mathbf{s}(t), a(t))\right]=\min_\pi J_\alpha(\mathbf{s}, \pi). 
\end{equation}

\begin{lemma} The $\alpha$-discounted value function satisfy the following Bellman equation:
\begin{equation}
V_\alpha(\mathbf{s})=\min_{a\in\mathbb{A}}\{C(\mathbf{s}, a)+\alpha\sum_{\mathbf{s}'}V_\alpha(\mathbf{s}')\rm{Pr}(\mathbf{s}'|\mathbf{s}, a)\}.
\label{Bellmanequation1}
\end{equation}
\end{lemma}

\hspace{0.2cm}\textit{Proof Sketch: }The main idea is to show there is a weight function $w(\mathbf{s}):\mathbb{S}\rightarrow[1, \infty)$ such that the $w$-norm of the value function $\Vert V_\alpha(\mathbf{s})\Vert=\sup_{\mathbf{s}\in\mathbb{S}}\frac{V_\alpha(\mathbf{s})}{w(\mathbf{s})}$ is bounded. The proof is similar to \cite{hsu17isit} and is provided in our online report due to page limitations.

Based on the Bellman equation, we can apply a policy iteration to obtain the value function $V_\alpha(\mathbf{s})$. The computational complexity of which can be reduced by utilizing the structure of the optimum policy. Next, we will first exploit the switching structure of the optimal policy $\pi_\alpha^*$ in Section III-A, and then propose a structural finite state policy iteration to approximate the optimal policy in Section III-B.
\subsection{Characterization of the Optimal Structure}

First we study the structure of the optimum policy $\pi_\alpha^*$ such that minimizes the discounted cost of MDP. We first present two lemmas, the proof of Lemma 3 them can be found in appendices.

\begin{lemma}
For fixed $\alpha$ and any starting state $\mathbf{s}$, the discounted value function $V_\alpha(\mathbf{s}+z\mathbf{e}_n)$ is a non-decreasing function of $z$, regardless of $n$.	
\end{lemma}

\begin{lemma}
	For any fixed $\alpha$, the discounted value function possess a submodularity characteristic. That is, for state $\mathbf{s}$ and $\forall i\neq j, z_i\geq 0, 0\leq z_j\leq s_j$:
	\begin{align}
	&V_\alpha(\mathbf{s}+z_i\mathbf{e}_i-z_j\mathbf{e}_j)-V_\alpha(\mathbf{s}-z_j\mathbf{e}_j)\nonumber\\
	\geq& V_\alpha(\mathbf{s}+z_i\mathbf{e}_i)-V_\alpha(\mathbf{s}).\end{align}
\end{lemma}

Based on the two lemmas, we will obtain the following theorem on the structure of $\pi_{\alpha}^*$:
\begin{theorem}
	The optimum policy $\pi_{\alpha}^*$ possesses a switching structure. That is, if for state $\mathbf{s}$ policy $\pi_{\alpha}^*$ chooses action $n$, then policy $\pi_{\alpha}^*$ chooses action $n$ at state $\mathbf{s}+z\mathbf{e}_n, \forall z\in\mathbb{N}$, where $\mathbf{e}_n$ is the unit vector with the $n^{\text{th}}$ component being 1 and the remaining elements being 0.
\end{theorem}
\begin{IEEEproof}
The proof is based on $N=2$ for notation simplicity and can be generalized easily to $N>2$. Suppose it is optimal to schedule user $1$ at state $\mathbf{s}=[s_1, s_2]$ with discount factor $\alpha$, then we have $\mathbb{E}_{s_1', s_2'}\left[V_\alpha([s_1', s_2'])|[s_1, s_2], 1\right]\leq \mathbb{E}_{s_1', s_2'}\left[V_\alpha([s_1', s_2'])|[s_1, s_2], 2\right]$. Then let us compute and compare the expected value function by taking action $a=1$ and $a=2$ at state $[s_1+z, s_2]$,
\begin{align}
	&\mathbb{E}_{s_1', s_2'}[V_\alpha([s_1', s_2'])|[s_1+z, s_2], 1]\nonumber\\
	&-\mathbb{E}_{s_1', s_2'}[V_\alpha([s_1', s_2'])|[s_1+z, s_2], 2]\nonumber\\
	=&p_1((1-\lambda_1)V_\alpha([0, s_2+1])+\lambda_1V_\alpha([1, s_2+1]))\nonumber\\
	&+(1-p_1)V_\alpha([s_1+z+1, s_2+1])\nonumber\\
	&-p_2((1-\lambda_2)V_\alpha([s_1+z+1, 0])+\lambda_2V_\alpha([s_1+z+1, 1]))\nonumber\\
	&-(1-p_2)V_\alpha([s_1+z+1, s_2+1])\nonumber\\
	=&p_1((1-\lambda_1)V_\alpha([0, s_2+1])+\lambda_1V_\alpha([1, s_2+1]))\nonumber\\
	&+(1-p_1)V_\alpha([s_1+1, s_2+1])\nonumber\\
	&-(1-p_1)V_\alpha([s_1+1, s_2+1])\nonumber\\
	&+(1-p_1)V_\alpha([s_1+z+1, s_2+1])\nonumber\\
	&-p_2((1-\lambda_2)V_\alpha([s_1+z+1, 0])+\lambda_2V_\alpha([s_1+z+1, 1]))\nonumber\\
	&-(1-p_2)V_\alpha([s_1+z+1, s_2+1])\nonumber\\
	\overset{(a)}{\leq}&p_2((1-\lambda_2)V_\alpha([s_1+1, 0])+\lambda_2V_\alpha([s_1+1, 1]))\nonumber\\
	&+(1-p_2)V_\alpha([s_1+1, s_2+1])\nonumber\\
	&+(1-p_1)(V_\alpha([s_1+z+1, s_2+1])-V_\alpha([s_1+1, s_2+1]))\nonumber\\
	&-p_2((1-\lambda_2)V_\alpha([s_1+z+1, 0])+\lambda_2V_\alpha([s_1+z+1, 1]))\nonumber\\
	&-(1-p_2)V_\alpha([s_1+z+1, s_2+1])\nonumber\\
	=&p_2(1-\lambda_2)(V_\alpha([s_1+1, 0])-V_\alpha([s_1+z+1, 0])\nonumber\\
	&-V_\alpha([s_1+1, s_2+1])+V_\alpha([s_1+z+1, s_2+1]))\nonumber\\
	&+p_2\lambda_2(V_\alpha([s_1+1, 1])-V_\alpha([s_1+z+1, 1])\nonumber\\
	&-V_\alpha([s_1+1, s_2+1])+V_\alpha([s_1+z+1, s_2+1]))\nonumber\\
	&-p_1(V_\alpha([s_1+z+1, s_2+1])-V_\alpha([s_1+1, s_2+1]))
	\overset{(b)}{\leq}0, \label{eq:diff}
\end{align}
where inequality (a) is obtained because it is optimum to broadcast source 1 at state $[s_1, s_2]$, which implies: \[\mathbb{E}_{s_1', s_2'}[V_\alpha([s_1', s_2'])|[s_1, s_2], 1]\leq\mathbb{E}_{s_1', s_2'}[V_\alpha([s_1', s_2'])|[s_1, s_2], 2],\] and (b) is obtained because of submodularity and monotonic. The above inequality Eq.~\eqref{eq:diff} implies, the optimum choice at state $[s_1+z, s_2]$ is selecting source $1$, which is the optimum action at state $[s_1, s_2]$. The switching structure is hence verified.
\end{IEEEproof}

\subsection{Relative Policy Iteration through Finite-State Approximation}
MDP problems with countable finite states can be solved by policy iteration or value iteration. To deal with the infinite state space in our problem, we approximate the whole countable space, i.e., the AoS for each source, by setting an upper bound of AoS $S_n^{\text{max}}$ for each of them. This approximation is reasonable since the probability of consecutive packet-loss vanishes exponentially with the number of consecutive transmission slots. By letting $S_n^{\text{max}}$ goes to infinity for all $n$, the optimal structure will converge to the original problem.

Denote $x_n^{(m)}(t)$ be the truncated AoS of source $n$ when $S_n^{\max}=m$, i.e., $x_n^{(m)}(t)=\min\{s_n(t), m\}$. With such approximation, by choosing different upper bound $m$, we can obtain a class of approximate MDP problems, where each problem differs from the primal problems with:
\begin{itemize}
	\item \textbf{State space: }We substitute the state $\mathbf{s}(t)$ by the truncated AoS $\mathbf{x}^{(m)}(t)=[x_1^{(m)}(t), x_2^{(m)}(t), \cdots, x_N^{(m)}(t)]^T$.
	
	\item \textbf{Transition probabilities: }The transition probability changes in accordance with the action space, let $\text{Pr}({\mathbf{x}^{(m)}}'|\mathbf{x}^{(m)}, a)$ be the transition probability from state $\mathbf{x}(t)=\mathbf{x}^{(m)}$ to $\mathbf{x}(t+1)={\mathbf{x}^{(m)}}'$ with the dynamic being
\end{itemize}
\begin{equation}
	\text{Pr}({\mathbf{x}^{(m)}}'|\mathbf{x}^{(m)}, a)=\prod_{n=1}^N\text{Pr}({x_n^{(m)}}'|x_n^{(m)}, a).
\end{equation}
It should be noted that $\text{Pr}({x_n^{(m)}}'|x_n^{(m)}, a)$ is the same as $\text{Pr}(s_n'|s_n, a)$ except:
\begin{equation}\begin{split}\text{Pr}({x_n^{(m)}}'|x_n^{(m)}, a)=\begin{cases}
1,&{x_n^{(m)}}'=x_n^{(m)}=m, a\neq n;\\
1-p_n,&{x_n^{(m)}}'=x_n^{(m)}=m, a=n.
\end{cases}\end{split}\end{equation}

Then for a given upper bound $m$, we can obtain an optimal deterministic policy by relative policy iteration. We choose the initial policy $\pi^{(0)}(\mathbf{x})=\arg_nx_n$, i.e., the greedy policy that schedules user with the largest AoS. Then given policy $\pi^{(k)}(\mathbf{x})$ and value function $V_\alpha^{(k)}(\mathbf{x})$, policy $\pi^{(k+1)}(\mathbf{x})$ and the value function $V_\alpha^{(k+1)}(\mathbf{x})$ in the $(k+1)^{\text{th}}$ iteration can be obtained through iteration. Considering the switching structure, once $\pi^{(k+1)}(\mathbf{x})=a$ is obtained, it can be concluded then for any $z\geq 0$, $\pi^{(k+1)}(\mathbf{x}+z\mathbf{e}_a)=a$. The policy $\pi^{(k)}(\mathbf{x})$ and value $V_\alpha^{(k)}(\mathbf{x})$ will finally converge when $k$ increases. Algorithm flowchart is provided below.
	\begin{algorithm}[h]
	\caption{Relative policy iteration based on switching structure}
	\begin{algorithmic}[1]
		\STATE \textbf{initialization}: for each state $\mathbf{x}$, assign action $\pi^{(0)}(\mathbf{x})=\arg\max_nx_n$, the initial value of $V_\alpha^{(0)}(\mathbf{x})=\sum_{n=1}^Nx_n$.
		
		\REPEAT
		
		\STATE{$\pi^{(k+1)}(\mathbf{x})\leftarrow0, \forall \mathbf{x}$.}
		
		\FOR{$\mathbf{x}\in$ state space $\mathcal{X}$ and $\pi^{(k)}(x)=0$}
		
		\STATE{Policy selection $ \pi^{(k+1)}(\mathbf{x})\leftarrow a^*, \text{where }a^*=\arg\min_{a\in\mathcal{A}}\{C(\mathbf{x}, a)+\alpha \mathbb{E}_{\mathbf{x}'}[V_\alpha^{(k)}(\mathbf{s}')|\mathbf{s}, a]\}$.}
		
		\STATE{Policy evaluation $V_\alpha^{(k+1)}(\mathbf{x})\gets C(\mathbf{x}, a^*)+\alpha\mathbb{E}_{\mathbf{x}'}[V_\alpha^{(k)}(\mathbf{x}')|\mathbf{x}, a^*]$}
		
		\STATE{Assign $\pi^{(k+1)}(\mathbf{x}+z\mathbf{e}_a)\gets a^*$ and $V_\alpha^{\text{tmp}}(\mathbf{x+z\mathbf{e}_a})\gets C(\mathbf{x}, a^*)+\alpha\mathbb{E}_{\mathbf{x}'}[V_\alpha^{(k)}(\mathbf{x}')|\mathbf{x}+z\mathbf{e}_a, a^*]$}
		
		\ENDFOR
		
		\STATE{$V_\alpha^{(k+1)}(\mathbf{x})\leftarrow V_\alpha^{\text{tmp}}(\mathbf{x})-V_\alpha^{\text{tmp}}(0), \text{for all } \mathbf{x}$.}
		
		\STATE{$k\leftarrow k+1$}
		
		\UNTIL{$\pi^{(k)}(\mathbf{x})=\pi^{(k-1)}(\mathbf{x}), \text{for all }\mathbf{x}$.}
		
	\end{algorithmic}
\end{algorithm}

After the iteration we can then obtain a stationary deterministic policy $\pi$. The MDP scheduling policy is obtained as follows: at each slot with state $\mathbf{s}(t)$, compute the corresponding virtual age $\mathbf{x}^{(m)}(t)$ and choose the corresponding action $a(t)=\pi(\mathbf{x}^{(m)}(t))$.

Notice that there are a total ${X_\text{max}}^N$ states, thus the computational complexity $\mathcal{O}({X_\text{max}}^N)$ grows exponentially, which makes the optimum policy impossible to obtain for large $N$. 
\section{Index-Based Heuristic}\label{sec:RMAB}

MDP solution is computationally demanding for a large number of access users known as the curse of dimension. To reduce computational complexity, we propose a simple index-based heuristic policy based on restless multi-arm bandit (RMAB) \cite{gittins2011multi}. 

The $N$ users can be viewed as arms, the state of user $n$ at the beginning of slot $t$ is the corresponding AoS $s_n(t)$. For the sake of simplicity, we define the bandit $n$ to be active if source $n$ is broadcasted, and the bandit is passive if source $n$ is not. In each slot, the BS activates one arm and sends update information, while the remaining arms remain passive. The AoS for each user $n$ depends only on its past AoS $s_n(t-1)$ and the scheduling decision $u_n(t-1)$. Hence the AoS evolves as restless bandit based on current action and its current AoS. 

To solve RMAB problem, a low complexity heuristic index policy is proposed by Whittle \cite{weber1990index}, it can approach asymptotic optimal performance compared with the MDP solution under certain scenarios. In this section, we first prove the indexability of the problem. Then, we obtain the closed form solution of Whittle's index and provide the scheduling policy.
\subsection{Decoupled Sub-Problem}
To adopt the Whittle's index, let us first relax the transmission constraints in each time slot into a time-average transmission constraint, \[\mathbb{E}_\pi\left[\frac{1}{T}\sum_{t=1}^T\sum_{n=1}^Nu_n(t)\right]\leq 1.\] 

Let $W\geq 0$ be the Lagrange multiplier and place the relaxed bandwidth constraint into the objective function, we have the following minimization problem:
	\begin{equation}
	\text{minimize }\lim_{T\rightarrow\infty}\mathbb{E}_\pi\left[\frac{1}{NT}\sum_{t=1}^T\sum_{n=1}^N\left(s_n(t)+Wu_n(t)-\frac{W}{N}\right)\right].
	\label{eq:banditopt}
	\end{equation}

For fixed $W$, the above relaxed optimization problem can be decoupled into $N$ subproblems and solved separately. Thus, we omit the subscript $n$ of each user henceforth. Given $W$, the goal of each decoupled optimization problem is to derive an optimum activation strategy $\mu_W$ such that the following total cost can be minimized: \begin{equation}
\mu_W=\arg_{\pi\in\Pi_{NA}}\lim_{T\rightarrow\infty}\frac{1}{T}\mathbb{E}_\mu\left[\sum_{t=1}^{T}\left(s(t)+Wu(t)\right)\right].\label{bandit}
\end{equation}The non-negative multiplier $W\geq 0$ can be viewed as an extra cost of being active and the optimum strategy should achieve a balance between activation cost and the cost incurred by AoS.

Each of the $N$ subproblems can be formulated into an MDP, each element is explained as follows:
\begin{itemize}
\item \textbf{State space:} The state at time $t$ is the current AoS of the corresponding user $s(t)\in\mathbb{N}$, which is countable infinite because of possible transmission failures.

\item \textbf{Action space:} There are two possible actions at each time slot, either choose the bandit to send updates $a(t)=1$ or remain idle $a(t)=0$. It should be noted here that the action $a(t)$ here is different to the scheduling action $u(t)$, which has strict interference constraint.

\item \textbf{Transition probability:} The state evolves with the action following Eq.~\eqref{stateevolution}. Let $\text{Pr}(s'|s, a)$ be the transition probability from state $s(t)=s$ to $s(t+1)=s'$ by taking action $a(t)=a$ at slot $t$, then:
\begin{equation}
\begin{split}
	\text{Pr}(s'|s, a)=\begin{cases}
	\lambda,&s'=1, s=0, a=0,1;\\
	1-\lambda,&s'=0, s=0, a=0,1;\\
	p\lambda,&s'=1, s\neq 0, a=1;\\
	p(1-\lambda),&s'=0, s\neq 0, a=1;\\
	1-p,&s'=s+1, s\neq 0, a=1;\\
	1,&s'=s+1, s\neq 0, a=0;\\
	0,&\text{otherwise}.
	\end{cases}
\end{split}
\label{banditprobability}
\end{equation}
\item \textbf{One-step cost: }For fixed $W$, the one step cost of state $s(t)$ by taking action $a(t)$ is defined as total cost increment at slot $t$, which consists of both the current AoS and the extra cost of being active:
\begin{equation}
C(s(t), a(t))=s(t)+Wa(t).
\end{equation}

\end{itemize}

According to \cite{Bertsekas_DP}, a stationary deterministic policy exists to minimize the average cost over infinite horizon. Next we will study the structure of such policy $\mu_W: s(t)\rightarrow a(t)$ to prove indexablility of the bandit.
\subsection{Proof of Indexability}
First, let us provide the formal definition of indexablility. 
\begin{definition}
	According to \cite{gittins2011multi}, let $\Omega_W$ be the set of states where the optimum strategy $\mu_W$ takes a passive action. A bandit is indexable if the passive set $\Omega_W$ increases monotonically with multiplier $W$, i.e., $\Omega_W\subset\Omega_{W'}, \forall W\geq W'$. 
\end{definition}

The proof for indexability can be divided into two parts. First we will show that the optimum policy to the above MDP possesses a threshold structure. Next, we derive the optimum threshold for fixed Lagrange multiplier $W$ and proves that it is monotonic increasing.

The threshold structure of the optimum stationary deterministic policy is obtained by investigating policies to minimize the $\alpha$-discounted cost over infinite horizon. For fixed Lagrange multiplier $W$, denote $J_{\alpha, W}(s, \mu)$ to be the $\alpha$-discounted over infinite horizon starting from initial state $s(1)=s$: 
\[J_{\alpha, W}(s, \mu)=\lim_{T\rightarrow\infty}\sup \mathbb{E}_\mu\left[\sum_{t=1}^T\alpha^{t-1}C(s(t), a(t))\right].\]

Let $\mu_{\alpha, W}=\arg\min_{\mu\in\Pi_{NA}}J_{\alpha, W}(s, \mu)$ be the optimum policy that minimizes the $\alpha$-discounted cost and let $V_{\alpha, W}(s)=\min_{\mu}J_\alpha(s, \mu)=J_{\alpha, W}(s, \mu_{\alpha, W})$. The value function $V_\alpha(s)$ satisfies the following Bellman equation:
\begin{equation}
\begin{split}
V_{\alpha, W}(s)=\min_{a\in\mathbb{A}}\{C(s, a)
+\alpha\sum_{s'}V_{\alpha, W}(s')\text{Pr}(s'|s, a)\}.
\end{split}
\label{Bellman}
\end{equation}
\begin{lemma}
	For $W>0$, the value function $V_{\alpha, W}(\cdot)$ increases monotonically.
\end{lemma}

The proof of Lemma 4 is provided in the appendix. Recall that $\mu_W$ is the optimum policy that minimizes the average cost Eq.~\eqref{eq:bandit}. Lemma 4 implies the following theorem about $\mu_W$:
\begin{theorem}
	The optimal policy $\mu_W$ to minimize the average cost over infinite horizon \eqref{bandit} has a threshold structure. Let $\mu_W(s)$ be the action policy $\mu_W$ takes when the AoS equals $s$. If at state $s$, it is optimal to keep the bandit idle, then for all $s'<s$ it is optimal to keep the bandit idle, i.e., $\mu_W(s')=0, \forall s'<s$; otherwise, if it is optimal to activate the bandit at state $s$, then for states $s+1, s+2, \cdots$, the optimal strategy $\mu_W$ is to activate the bandit, i.e., $\mu_W(s')=1, \forall s'>s$.
\end{theorem}
\begin{IEEEproof}
We will prove by investigating the optimal policy $\mu_{\alpha, W}$ that minimizes the $\alpha$-discounted cost. The decision $a(t)$ is chosen according to the Bellman equation \eqref{Bellman}, the condition for bandit to be active is:
\begin{equation}
W+\alpha p(\lambda V_{\alpha, W}(1)+(1-\lambda)V_{\alpha, W}(0))\leq \alpha pV_{\alpha, W}(s+1).
\end{equation}

Suppose at state $s$, the above inequality is satisfied and the optimum strategy $\mu_\alpha$ to minimize the $\alpha$-discounted cost is to choose the bandit to be active, i.e., $a=1$. According to the monotonic characteristic, for all states $s'$ satisfy $s'>s$, 
\begin{align}
&W+\alpha p(\lambda V_{\alpha, W}(1)+(1-\lambda)V_{\alpha, W}(0))\nonumber\\
\leq&\alpha pV_{\alpha, W}(s)\leq  \alpha pV_{\alpha, W}(s').\nonumber
\end{align} 

Hence for state $s'\geq s$, the optimum policy $\mu_{\alpha, W}$ will choose the bandit to be active, i.e., $\mu_{\alpha, W}(s')=1, \forall s'>s$. By taking the conversion of inequality, in the same way, we can obtain that the if the optimum policy $\mu_{\alpha, W}$ is to remain passive at state $s$, then for all states satisfy $s'\leq s$, the optimum policy to minimize $\alpha$-discounted cost is to remain passive. The threshold policy holds for all $\alpha\in(0, 1)$. By taking $\alpha\rightarrow 1$, this provides insight that the optimum policy $\mu_W$ has a threshold structure.
\end{IEEEproof}

The indexability is then proved by showing the activation threshold increases with $W$. To compute the activation threshold, first we compute the average cost for fixed Lagrange cost is threshold policy $\tau$ is employed, the proof can be found in Appendix \ref{Sec:C1}:

\begin{corollary}
	Denote $F(\tau, W)$ to be the average cost for fixed $W$ if threshold policy $\tau$ is employed, i.e., the bandit will be active for state $s\geq\tau$ and be passive for state $s<\tau$. Then,
		\begin{equation}
F(\tau,W)=\frac{\tau(\tau-1)}{2}\xi_1^{(\tau)}+\frac{\xi_1^{(\tau)}}{p}(\frac{1}{p}-1)+\frac{\xi_1^{(\tau)}}{p}(\tau+W),
\end{equation}
where $\xi_s^{(\tau)}$ denotes the steady state distribution if bandit is in state $s$ by applying threshold policy $\tau$ and typically,
\[\xi_1^{(\tau)}=1/\left(\frac{1-\lambda}{\lambda}+\tau+\frac{1}{p}-1\right).\]
\end{corollary}
Next, we derive the optimum activating threshold $\tau_{opt}(W)$ for given $W$ by examining the value of $F(\tau, W)$. The optimum value should satisfy $F(\tau_{opt}+1,W)\geq F(\tau_{opt},W)$ and $F(\tau_{opt}-1,W)\geq F(\tau_{opt},W)$. The following corollary provides the closed form expression of the threshold, further derivations are provided in Appendix \ref{Sec:C2}:
\begin{corollary}
	For given Lagrange multiplier $W$, the optimum activating threshold can be computed as follows: 
	\begin{align}
	&\tau_{opt}=\lfloor\left(\frac{5}{2}-\frac{1}{p}-\frac{1}{\lambda}\right)\label{thresholdtau}\\
	&\hspace{0.5cm}+\sqrt{\left(\frac{5}{2}-\frac{1}{p}-\frac{1}{\lambda}\right)^2+2(\frac{W}{p}-\frac{1-\lambda}{\lambda}\frac{1-p}{p})+2\frac{1-p}{p}}\rfloor.\nonumber\footnote{For the special case $\lambda=1$, the optimum threshold then becomes $\tau_{opt}=\lfloor\frac{3}{2}-\frac{1}{p}+\sqrt{(\frac{3}{2}-\frac{1}{p})^2+2\frac{W}{p}+2\frac{1-p}{p}}\rfloor=\lfloor\frac{3}{2}-\frac{1}{p}+\sqrt{(\frac{1}{2}-\frac{1}{p})^2+2\frac{W}{p}}\rfloor$. This special case is equivalent to the optimum threshold of AoI problem, \cite[Eq.~(54)]{igor_ton_2018} with $T=1$ and $\alpha=1$.}
	\end{align}
\end{corollary}

Notice that the scheduling threshold $\tau_{opt}$ is an increasing function of $W$, suggesting that the passive set increases monotonically with $W$. Especially when $W=0$ the threshold equals $0$, which suggests the passive set is $\emptyset$. Thus the indexability of the bandit is proved.
	
\subsection{Derivation of the Whittle's Index}
The Whittle's index $I(s)$ measures how rewarding it is if the bandit at state $s$ is activated. By definition, it is the extra cost that makes action $a=1$ and $a=0$ for states $s$ equally desirable \cite{weber1990index}. Denote $\phi(\tau)$ to be the average activation probability over infinite horizon if threshold policy $\tau$ is applied. According to the threshold structure, $\phi(\tau)$ equals the time proportion spent on updating the bandit, which equals the total probability that the bandit is in state $s\gamma \tau$ and can be computed by:
\begin{align}
	\phi(\tau)&=\sum_{s=\tau}^\infty \xi_s^{(\tau)}=\sum_{s=\tau}^\infty \xi_1^{(\tau)}(1-p)^{s-\tau}=\frac{\xi_1^{(\tau)}}{p}.
\end{align}

According to \cite[Eq.~6.11]{gittins2011multi}, the Whittle's index can be computed as follows:
\begin{equation}
	I(s)=\frac{F(s+1, 0)-F(s, 0)}{\phi(s)-\phi(s+1)}=\frac{p(F(s+1,0)-F(s,0))}{\xi_1^{(s)}-\xi_1^{(s+1)}}.\footnote{When $\lambda=1$, we have $\xi_1^{(s)}=\frac{1}{s+\frac{1}{p}-1}$ and $F(s+1, 0)-F(s, 0)=\left(\frac{s(s-1)}{2}+\frac{1}{p^2}-\frac{1}{p}+\frac{s}{p}\right)(\xi_1^{(s+1)}-\xi_1^{(s)})+(s+\frac{1}{p})\xi_1^{(s+1)}$. The Whittle's index according to our derivations is $I(s)=-p\left(\frac{s(s-1)}{2}+\frac{1}{p^2}-\frac{1}{p}+\frac{s}{p}\right)+p(s+\frac{1}{p})(s+\frac{1}{p}-1)=\frac{ps}{2}(\tau+\frac{2-p}{p})$, which is exactly equivalent to \cite[Eq.~(56)]{igor_ton_2018} with $T=1$ and $\alpha=1$.}
\end{equation}

\subsection{Index-Based Scheduling Algorithm}
We will provide a low-complexity scheduling algorithm in this part based on the derived index. At the beginning of each time slot, the BS observes current AoS of each source $s_n(t)$ and computes the Whittle's index for each user $I_n(s_n(t))$. Then, broadcast the corresponding message of user $n$ with the highest $I_n(s_n(t))$, with ties broke arbitrarily. Since the index is computed separately for each user, the computational complexity is $\mathcal{O}(N)$ and can almost be neglected. 

The scheduling policy based on Whittle's index can be easily generated to constraint where no more than $M$ users can be scheduled in each time slot. That is, after computing the Whittle's index of each user at time slot $t$, the BS selects $M$ users with the largest indexes and broadcast the corresponding update packet. Moreover, according to \cite{weber1990index}, the Whittle's index policy is shown to be asymptotic optimum in most scenarios. That is, let the number of users $N\rightarrow\infty$ while $M/N$ keeps a constant, the performance gap between Whittle's index policy and the optimum policy vanishes. 

\section{Numerical Simulations}
In this section, the performance of the proposed scheduling strategies are evaluated in terms of the expected average AoS over the entire network. We compare four scheduling strategies: 1) The greedy arrival aware policy that schedules to transmit undelivered packet to user with the largest AoS. 2) AoI minimization policy proposed in \cite{sun2019optimizing}. 3) The Markov decision process in Section~\ref{sec:MDP}. 4) The Whittle's index policy in Section~\ref{sec:RMAB}. Define the total packet arriving rate over the entire network to be $\lambda_{total}=\sum_{n=1}^N\lambda_n$. The expected average AoS is computed by taking the average AoS evolution over $T$ time slots such that each user is selected for transmission larger than a consecutive of $10^4$ slots.

%
\begin{figure}[h]
	\centering
	\includegraphics[width=.45\textwidth]{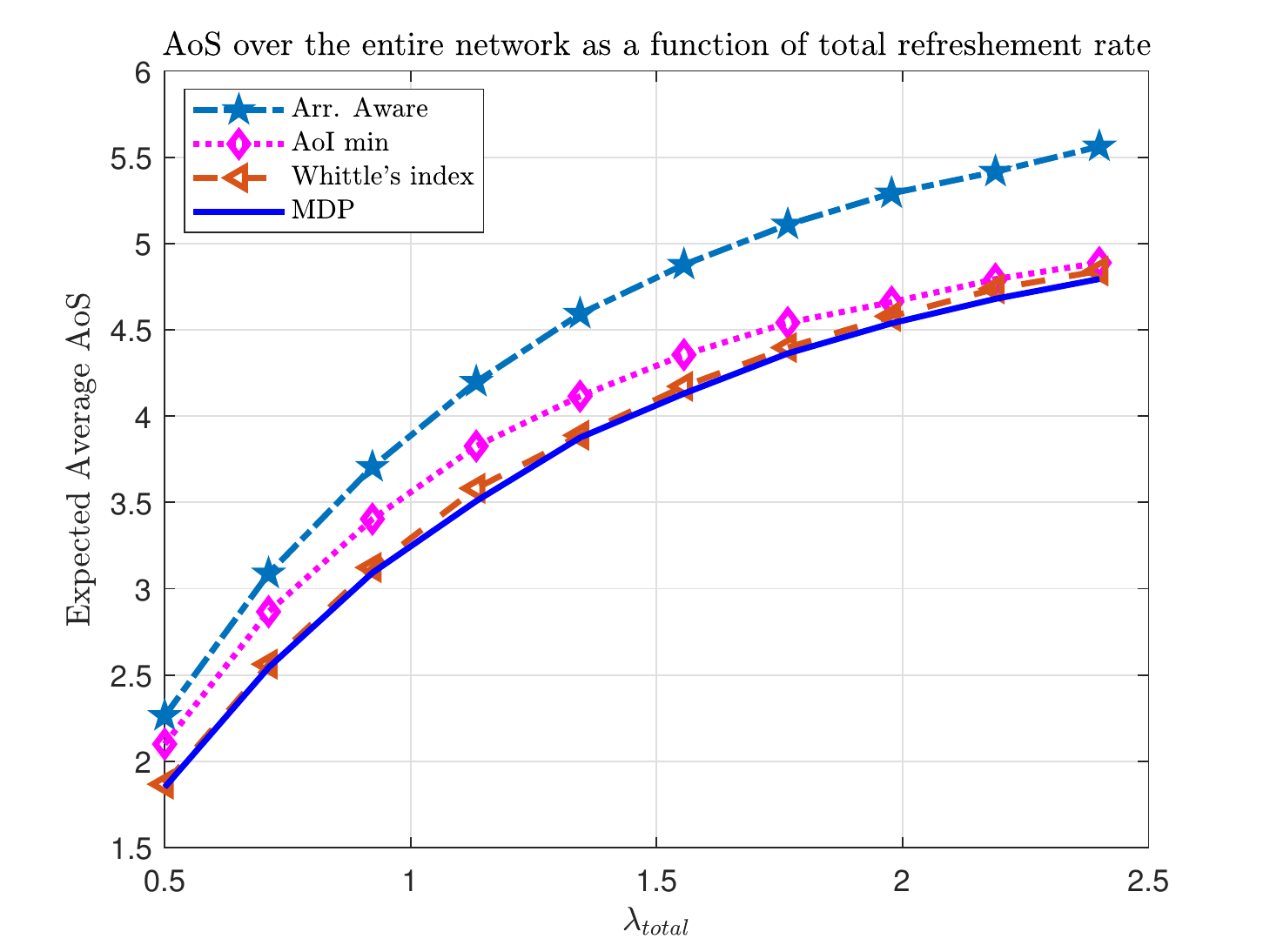}
	\caption{Simulation results of expected AoS for a three user broadcasting network with $\lambda=[0.3, 0.4, 0.3]\lambda_{total}$ and $p=[0.2, 0.55, 0.9]$}
	\label{RMABCmp}
\end{figure}	
\begin{figure}[h]
	\centering
	\includegraphics[width=.45\textwidth]{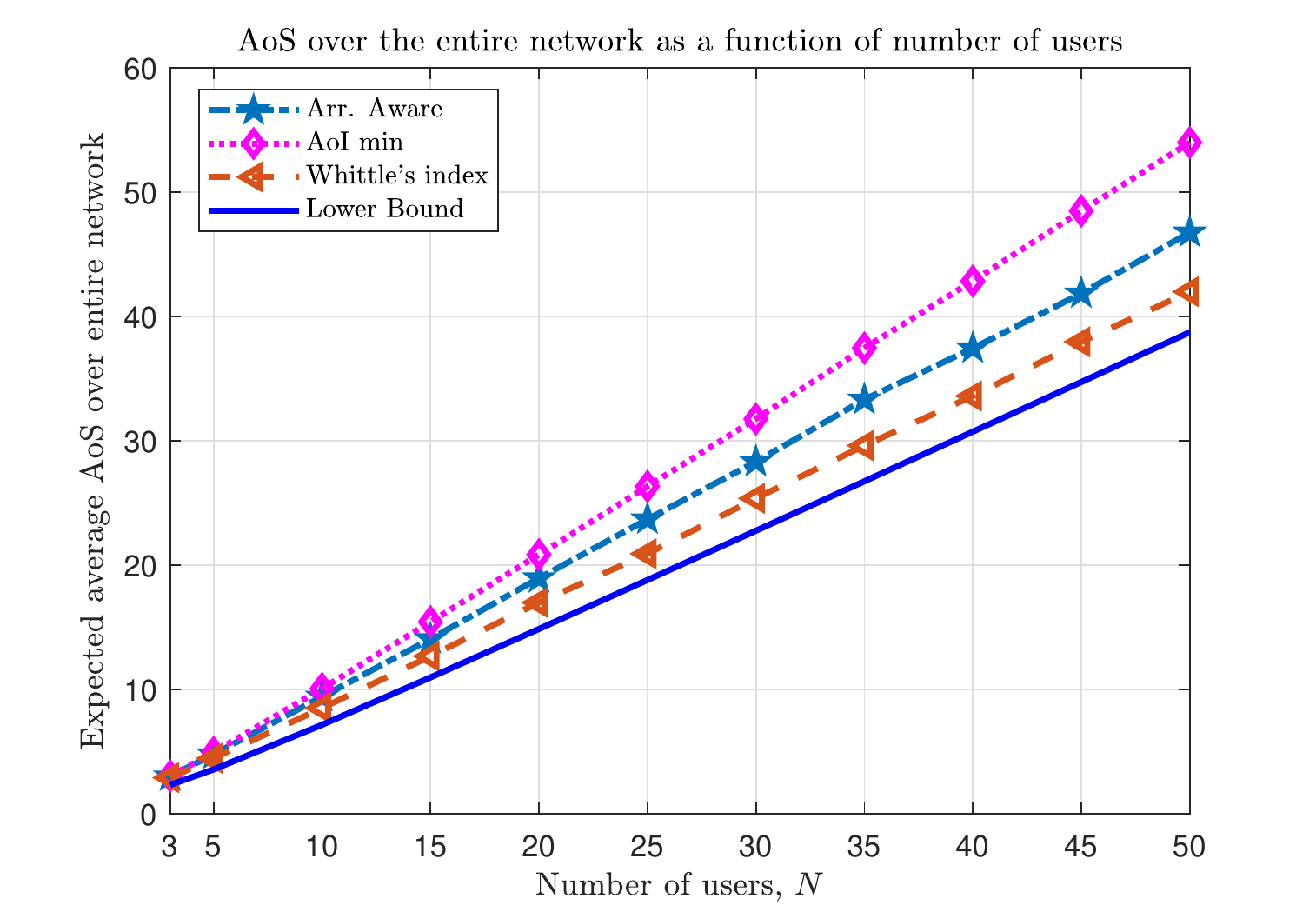}
	\caption{Simulation results of expected AoS as the number of users, the total packet arrival rate over the entire network $\lambda_{total}=2$ and the packet arrival probability for each user is $\lambda_n=\frac{2n}{N(N+1)}\lambda_{total}$, $p_n=n/N$.}
	\label{simu3}
\end{figure}
In Fig.~\ref{RMABCmp}, we consider a three user broadcast network with arriving rate $\lambda=[0.3, 0.4, 0.3]\lambda_{total}$ and success transmission probability $p=[0.2, 0.55, 0.9]$. The threshold $m$ for computing the truncated MDP solution is set to be $m=20$. Fig. \ref{simu3} study the AoS performance for networks with more users. The parameter $\lambda_{total}=2$, the packet arrival probability for each user is $\lambda_n=\frac{2n}{N(N+1)}\lambda_{total}$ and $p_n=\frac{n}{N}$. Due to the computational complexity caused by the curse of dimension, we display the derived lower bound instead of the MDP policy in Fig.~\ref{simu3}. In Fig. \ref{RMABCmp}, the proposed index based scheduling algorithm achieves compatible performance with the MDP policy. In Fig.~\ref{simu3},
the performance of the proposed index policy is close to the theoretic lower bound. When the number of users $N$ increases, following arrival aware strategy is far from the proposed index policy.

From our analysis in Section II, AoI and AoS are equivalent when $\lambda=1$. In Fig.~\ref{RMABCmp}, when $\lambda_{total}\rightarrow 2.4$, the arrival rate $\lambda_n$ for each user is close to 1, indicating that update packets will arrive in nearly every time slot. AoI minimization policy tends to show similar performance as the AoS minimization policies. When the update packet arrival rate $\lambda_n$ is much less than 1, or the number of users in networks increases, AoI minimization policy will lead to significantly higher AoS performance compared with the proposed index policy and may even be worse than the greedy arrival aware policy. This phenomenon verifies our analysis that AoS and AoI are metrics with different physical meanings, a good AoI performance cannot guarantee a good AoS performance.

\begin{figure*}[h]
	\centering
	\includegraphics[width=.7\textwidth]{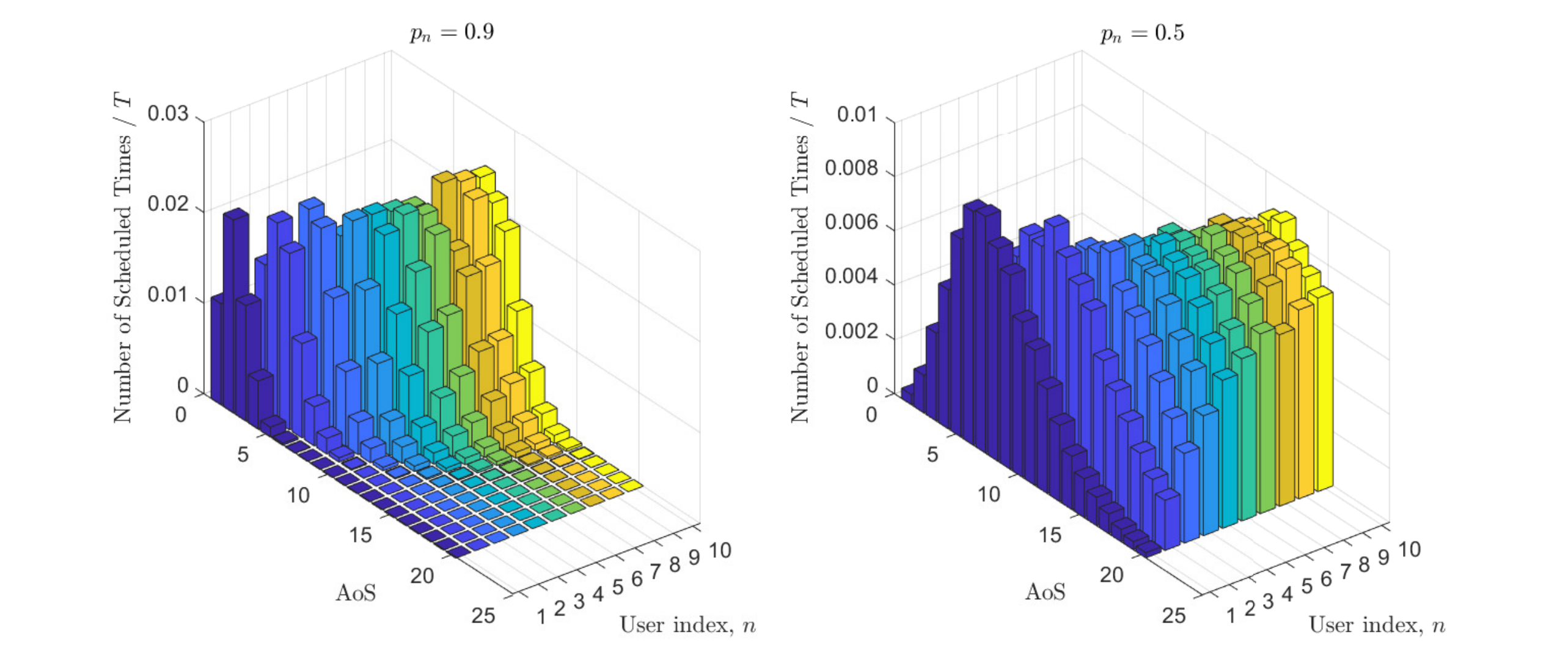}
	\caption{Simulations of the time proportions spent on scheduling each user with different AoS, i.e., the number of scheduling times divide the total number of slots $T$ for a network with $N=10$ users, with $\lambda_n=0.5n/N$. The transmission success probability is $p_n=0.9,\forall n$ on the left and $p_n=0.5$ on the right. }
	\label{fig:decisionlambda}
\end{figure*}

\begin{figure*}[h]
	\centering
	\includegraphics[width=.8\textwidth]{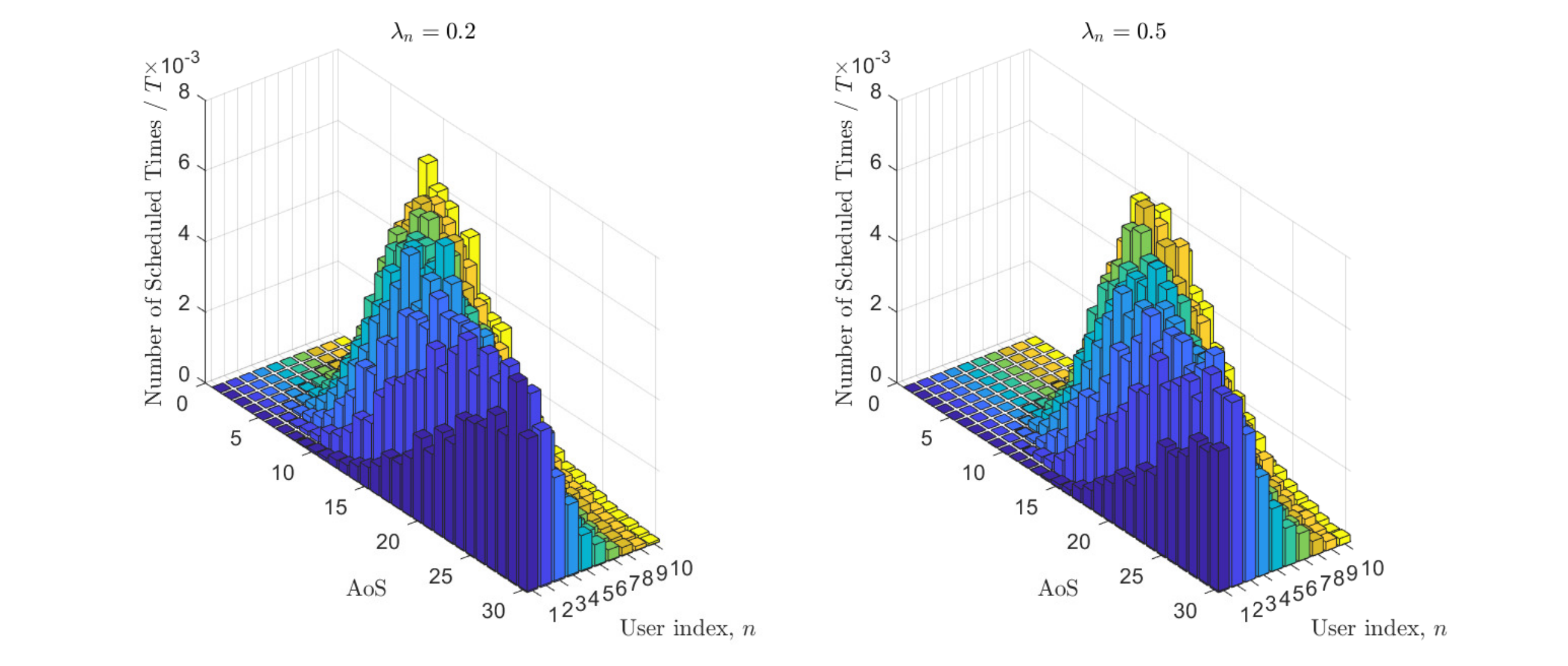}
	\caption{Simulations of the time proportions spent on scheduling each user with different AoS, i.e., the number of scheduling times divide the total number of slots $T$ for a network with $N=10$ users, with $p_n=n/N$. The random update probability in each slot is $\lambda_n=0.2\forall n$ on the left and $\lambda_n=0.5$ on the right. }
	\label{fig:decisionp}
\end{figure*}

To understand the design insight of how to minimize the average AoS over the entire network, we plot the time proportion spent on scheduling each user under different AoS, i.e., the number of scheduling times divide the total number of slots $T$. In Fig.~\ref{fig:decisionlambda}, we study a network with $N=10$ users, with packet arrival rate $\lambda_n=0.5n/N$ for each user. The transmission success probability is $p_n=0.9$ on the left and $p_n=0.5$ on the right for all $n$. In Fig.~\ref{fig:decisionp} we set $p_n=n/N$ for each user, and the update arrival probability in each slot is $\lambda_n=0.2$ on the left and $\lambda_n=0.5$ on the right. From simulations, users with smaller update arrival probability $\lambda_n$ and larger packet transmission success probability are more likely to be updated with smaller AoS.
\begin{section}{Conclusions}
In this paper, we treated a broadcast network with a BS sending random updates to interested users over unreliable wireless channels. We measure data freshness by \emph{Age of Synchronization}. We propose two scheduling algorithms based on Markov decision process (MDP) and the restless multi-arm bandit (RMAB) to minimize AoS. Simulation results show that the proposed index policy achieves comparable performance with the MDP and approaches the theoretic lower bound. Moreover, our work verifies that AoS and AoI are different concepts, policy mismatch will lead to bad AoS performances. To guarantee a good AoS performance, scheduling policies should ensure that users with smaller random update probabilities and larger success probabilities are updated at smaller AoS.

The wireless broadcast model is a simplified one in our work. In the future, we will study scheduling policies to minimize AoS in broadcast network with time-varying channels like \cite{timevarying}. Scheduling and interference alignment co-design in a network with multiple base stations will also be an interesting problem. 
\end{section}
\section*{Acknowledgment}
	The authors would like to thank Mr. Jingzhou Sun, Prof. Zhiyuan Jiang, Prof. Sheng Zhou and Prof. Zhisheng Niu for a preprint of the manuscript \cite{sun2019optimizing} and fruitful discussions. The authors would like to thank Prof. Roy Yates for suggestions during the ISIT and are also grateful to the anonymous reviewers for valuable suggestions that improve the presentation of the manuscript.
\appendices
\begin{section}{Proof of Theorem 1}\label{sec:thm1proof}
\begin{IEEEproof}Let $\pi\in\Pi_{NA}$ be a feasible scheduling policy satisfying the interference constraint. Since there will be no AoS decrease to broadcast source $n$ when the AoS of user $n$ equals 0, to formulate the AoS lower bound, we assume policy $\pi$ broadcast source $n$ at time $t$ only if $s_n(t)>0$. Following policy $\pi$, a sample path of AoS, denote by $\omega$ is obtained. Suppose up to slot $T$, the BS broadcast update packets of source $n$ for a total of $L_n^T$ times, and user $n$ receives $K_n^T$ packets successfully. Similar to the analysis in \cite{igor_ton_2018}, when $T\rightarrow\infty$, any strategy that transmits updates to user $n$ less than a fixed constant times will lead to an infinite average AoS and is thus far from optimum. We call such strategies to be "starving strategies". In the following analysis, we only focus on policy $\pi$ that belongs to "non-starving strategies", which implies the number of broadcasts of source $n$ goes to infinity when $T$ goes to infinity, i.e.,  \begin{equation}
	\lim_{T\rightarrow\infty}L_n^T=\infty, \text{w.p.1.}\label{eq:Tinf}
	\end{equation}
	Since each broadcast arrives at user $n$ with probability $p_n$, by the law of large numbers, we have: \begin{equation}\lim_{T\rightarrow\infty}\frac{K_n^T}{L_n^T}=p_n, \text{w.p.1}. \label{eq:pinf}
\end{equation} The above equations imply that the number of received packets about source $n$ by user $n$ goes to infinity with probability 1, i.e., \begin{equation}\lim_{T\rightarrow\infty}K_n^T=\infty, \text{w.p.1.}
	\end{equation}

	Suppose user $n$ receives the $i^{\text{th}}$ update packet about source $n$ at the end of slot $t_{n, i}$. Denote $\tau_{n, i}$ to be the inter-update interval of user $n$ between the receiving time-stamps of the $(i-1)^{\text{th}}$ and the $i^{\text{th}}$ update, which can be computed as follows:
	\begin{equation}
	\tau_{n, i}=t_{n, i}-t_{n, i-1}, \forall i\in\{1, 2, \cdots, K_n^T\}.
	\end{equation}
	Since all users are assumed to be synchronized initially, let $t_{n, 0}=0, \forall n$. To facilitate the AoS computation during slots $[t_{n, K_n^T}, T]$, we define $\tau_{n, K_n^T+1}$ to be:
	\begin{equation}
		\tau_{n, K_n^T+1}=T-t_{n, K_n^T}. 
	\end{equation}
	Then sum of sequence $\{\tau_{n, i}\}$ satisfies:
	\begin{equation}
	\sum_{i=1}^{K_n^T+1}\tau_{n, i}=T.
	\end{equation}
	
	According to the AoS evolution, if source $n$ has no updates after the latest update packet has been received by user $n$, the AoS $s_n(t)$  keeps zero. Denote $v_{n, i}$ to be the maximum number of consecutive slots that the AoS of user $n$ remains $0$ after the $i^{\text{th}}$ update packet has been received by user $n$, i.e., $s_n(t_{n, i}+j)=0, \forall j\in[1, v_{n,i}]$ and $s_n(t_{n, i}+v_{n, i}+1)=1$. The AoS of user $n$ will start from $1$ at the beginning of slot $t_{n, i}+v_{n, i}+1$ and increases linearly until the next update packet has been received by the end of slot $t_{n, i+1}$. For simplicity, we denote $w_{n, i}=s_n(t_{n, i+1})=t_{n,i+1}-t_{n,i}-v_{n,i}$ afterwards. The total AoS of user $n$ at the beginning of each slot between $[t_{n, i}+1, t_{n, i+1}]$ can be computed as follows:
	\begin{equation}
	\sum_{j=t_{n, i}+1}^{t_{n, i+1}}s_n(j)=\frac{s_n(t_{n, i+1})(s_n(t_{n, i+1})+1)}{2}=\frac{w_{n, i}^2}{2}+\frac{w_{n, i}}{2}.
	\label{intervalsum}
	\end{equation}
	
	Let $S_T(\omega)$ be the average AoS over slot $[1, T]$ of sample path $\omega$, which can be computed by:
	\begin{align}
	S_T(\omega)&=\frac{1}{NT}\sum_{n=1}^N\sum_{i=1}^{K_n^T+1}\left(\sum_{j=t_{n, i-1}+1}^{t_{n, i}}s_n(j)\right)\nonumber\\
	&=\frac{1}{N}\sum_{n=1}^N\left(\frac{K_n^T+1}{T}\frac{\sum_{i=1}^{K_n^T+1}\frac{w_{n, i}^2}{2}+\frac{w_{n, i}}{2}}{K_n^T+1}\right).
	\label{AoSsamplemean1}
	\end{align}
	Let $\mathbb{M}[\cdot]$ denote the sample mean of a set of variables and denote $\gamma_n=\frac{K_n^T+1}{T}$. Then $S_T(\omega)$ can be simplified and upper bound as follow:
	\begin{align}
S_T(\omega)&=\frac{1}{N}\sum_{n=1}^N\gamma_n\left(\frac{1}{2}\mathbb{M}[w_n^2]+\frac{1}{2}\mathbb{M}[w_n]\right)\nonumber\\
&\overset{(a)}{\geq}\frac{1}{N}\sum_{n=1}^N\gamma_n\left(\frac{1}{2}\mathbb{M}[w_n]^2+\frac{1}{2}\mathbb{M}[w_n]\right),\label{eq:LB1}
\end{align}	
where inequality (a) is obtained by the generalized mean inequality $\mathbb{M}[w_n^2]\geq\mathbb{M}[w_n]^2$. To further lower bound $S_T(\omega)$, in the following analysis, we first compute $\mathbb{M}[w_n]$ for each fixed $\gamma_n$, then figure out the constraints on values sequences $\{\gamma_n\}$ can take. Finally, searching for the lower bound of AoS can be formulated into an optimization problem, where the objective is to minimize the lower bound of $\lim_{T\rightarrow\infty}\mathbb{E}[S_T(\omega)]$ under constraints about $\{\gamma_n\}$.

	Given $\gamma_n$, the computation of $\mathbb{M}[w_n]$ is divided into two steps. First, we obtain $\mathbb{M}[w_n]+\mathbb{M}[v_n]$ and then we compute $\mathbb{M}[v_n]$. The sum of $\mathbb{M}[w_n]$ and $\mathbb{M}[v_n]$ is obtained as follows:
	\begin{align}
		&\mathbb{M}[w_n]+\mathbb{M}[v_n]\nonumber\\
		\overset{(a)}{=}&\frac{\sum_{i=1}^{K_n^T+1}w_{n, i}}{K_n^T+1}+\frac{\sum_{i=1}^{K_n^T+1}v_{n,i}}{K_n^T+1}\nonumber\\
		\overset{(b)}{=}&\frac{T}{K_n^T+1}=\frac{1}{\gamma_n},\label{eq:summean}
	\end{align}
	where equality (a) is obtained by definition of $\mathbb{M}[\cdot]$ and equality (b) is because $\tau_{n, i+1}=w_{n, i}+v_{n, i}$ and $T=\sum_{i=1}^{K_n^T+1}\tau_{n, i}$. 
%
	
	Since update packet arrives independently in each slot with probability $\lambda_n$, the consecutive number of slots that no update packets appear $v_{n, i}$ are i.i.d random variables with geometric distribution of coefficient $\lambda_n$. By the law of large numbers, 
	\begin{equation}
	\mathbb{M}[v_n]=\mathbb{E}[v_{n, i}]=\frac{1-\lambda_n}{\lambda_n}, \text{w.p.1.}
	\label{eq:vmean}
	\end{equation}
	
	Plugging Eq.~\eqref{eq:vmean} into the Eq.~\eqref{eq:summean}, the mean value $\mathbb{M}[w_n]$ can be written out as a function of $\gamma_n$:
	\begin{align}
	\mathbb{M}[w_n]=\frac{1}{\gamma_n}-\frac{1}{\lambda_n}+1, \text{w.p.1}.\label{eq:meanw}
	\end{align}
	
	The first constraint on $\gamma_n$ is obtained through the lower bound of $\mathbb{M}[w_n]$. Recall policy $\pi$ only broadcast updates of source $n$ when the AoS of user $n$ is no longer 0. Thus, $s_n(t_{n, i})\geq 1$, which implies $\mathbb{M}[w_n]\geq 1$. According to Eq.~\eqref{eq:meanw}, we have the following restrictions on $\gamma_n$:
	\begin{equation}
	\frac{1}{\gamma_n}-\frac{1}{\lambda_n}\geq 0\Rightarrow \gamma_n\leq \lambda_n,\text{w.p.1.}\label{eq:constraintw}
	\end{equation}
	
	The second constraint on $\{\gamma_n\}$ inherits from the interference constraint that no more than one source can be broadcasted in each slot. By summing up the interference constraint in each slot, i.e.,  Eq.~\eqref{eq:primalconstraint}, from $t=1$ to $T$, we can obtain the following inequality on sequence $\{L_n^T\}$, i.e.,
	\begin{equation}
	\sum_{n=1}^{N}L_n^T\leq T.
	\label{relaxcons}
	\end{equation}
	
	Recall that when $T\rightarrow\infty$, we have $\gamma_n=\lim_{T\rightarrow\infty}\frac{K_n^T+1}{T}=\lim_{T\rightarrow\infty}\frac{K_n^T}{T}$. By plugging Eq.~\eqref{eq:pinf}, i.e., the relationship between $K_n^T$ and $L_n^T$ into Eq.~\eqref{relaxcons}, we have: \begin{equation}
	\sum_{n=1}^N\frac{\gamma_n}{p_n}\leq 1.\label{eq:bandwidth}
	\end{equation}
	
	Thus, for any non-starving policy $\pi$ that satisfies the bandwidth constraint, the lower bound on its AoS performance can be computed by computing $\{\gamma_n\}$, and then place Eq.~\eqref{eq:meanw} into Eq.~\eqref{eq:LB1}. Based on the above analysis, let $\text{AoS}_{\text{LB}}$ be the average AoS lower bound over the entire network, searching for $\text{AoS}_{\text{LB}}$ can be formulated into the following optimization problem:
	\begin{subequations}
		\begin{align}
		&\text{AoS}_{\text{LB}}=\min_{\boldsymbol{\gamma}\geq 0}\frac{1}{N}\sum_{n=1}^N\gamma_n\left[\frac{1}{2}\left(\frac{1}{\gamma_n}-\frac{1}{\lambda_n}+1\right)^2\right.\nonumber\\
		&\hspace{4cm}\left. +\frac{1}{2}\left(\frac{1}{\gamma_n}-\frac{1}{\lambda_n}+1\right)\right],\label{Eq:AoSLB}
		\end{align}
		\begin{align}
		\text{s.t, }&\sum_{n=1}^N\frac{\gamma_n}{p_n}\leq 1,\\
		&\gamma_n \leq \lambda_n, \text{for all }n=1, 2, \cdots, N,
		\end{align}
	\end{subequations}
	For any policy $\pi$, we can conclude that the average AoI obtained over $T$ consecutive slots is larger than $\text{AoS}_{\text{LB}}$ with probability 1 when $T\rightarrow\infty$. We will then proceed to talk about how the solve the above optimization problem. 
	
	Notice that the objective function is convex with closed polygon constraint. For simplicity, denote $\boldsymbol{\gamma}=[\gamma_1, \cdots, \gamma_N]$ and $\boldsymbol{\nu}=[\nu_1, \cdots, \nu_N]$. We can write out the Lagrange function as follows: 
	\begin{align}\mathcal{L}(&\boldsymbol{\gamma}, \mu, \boldsymbol{\nu})=\nonumber\\
	&\frac{1}{N}\sum_{n=1}^N\gamma_n\left(\frac{1}{2}\left(\frac{1}{\gamma_n}-\frac{1-\lambda_n}{\lambda_n}\right)^2+\frac{1}{2}\left(\frac{1}{\gamma_n}-\frac{1-\lambda_n}{\lambda_n}\right)\right)\nonumber\\
	&+\mu\left(\sum_{n=1}^N\frac{\gamma_n}{p_n}-1\right)+\sum_{n=1}^N\nu_n(\gamma_n-\lambda_n),
	\end{align} where $\mu$ and $\boldsymbol{\nu}\geq 0$ are the Lagrange multipliers. According to the Karush-Kuhn-Tucker (KKT) conditions, when the function reaches its minimum, the condition $\nabla_{\boldsymbol{\gamma}}\mathcal{L}(\boldsymbol{\gamma}, \mu, \boldsymbol{\nu})=0$ holds for any $n$, i.e.,
	\begin{equation}
	\frac{1}{2N}\left[\left(\frac{1-\lambda_n}{\lambda_n}\right)^2-\left(\frac{1-\lambda_n}{\lambda_n}\right)\right]-\frac{1}{2N}\frac{1}{\gamma_n^2}+\frac{\mu}{p_n}+\nu_n=0.
	\label{KKT}
	\end{equation}
	Hence, the optimum $\gamma_n$ can be expressed as a function of $\mu$ and $\nu_n$:
	\begin{equation}
	\gamma_n=1/\sqrt{\left(\frac{1-\lambda_n}{\lambda_n}\right)^2-\frac{1-\lambda_n}{\lambda_n}+N\left(\frac{2\mu}{p_n}+2\nu_n\right)}.
	\end{equation}
	
	Notice that the primal constraint $\gamma_n\leq\lambda_n$ implies that $\nu_n=0, \text{ if }\gamma_n<\lambda_n$. Next, consider the Complete Slackness (CS) conditions:
	\begin{subequations}
	\begin{align}
	\mu\left(\sum_{n=1}^N\frac{\gamma_n}{p_n}-1\right)&=0,\\
	\nu_n(\gamma_n-\lambda_n)&=0, \forall n. 
	\end{align}
	\end{subequations} Then the optimum $\gamma_n^*$ can be computed as follows:
	\begin{equation}
	\gamma_n^*=\max\{1/\sqrt{\left(\frac{1-\lambda_n}{\lambda_n}\right)^2-\left(\frac{1-\lambda_n}{\lambda_n}\right)+\frac{2\mu^*N}{p_n}}, \lambda_n\},
	\end{equation}
	where $\mu^*$ is the Lagrange multiplier that keeps $\sum_{n=1}^N\frac{\gamma_n^*}{p_n}=1$.
	\end{IEEEproof}
\end{section}

\begin{section}{Proof of Lemma 3}
		To simplify the notations, we consider $N=2$ and denote $\mathbf{s}=[s_1, s_2]$ in the following discussion. The analysis can be generalized $N>2$. Similarly, we prove the submodularity investigating into the Bellman operator. Suppose $V_\alpha^{(k)}(\cdot)$ has the submodularity characteristic, we will then show that $V_\alpha^{(k+1)}(\cdot)$ obtained after the $(k+1)^{\text{th}}$ iteration possesses the same characteristic.
	
	With no loss of generality, assume $i=1, j=2$. By the submodularity of $V_\alpha^{(k)}(\cdot)$, we have
	\begin{align}&V_\alpha^{(k)}([s_1+z_1, s_2-z_2])+V_\alpha^{(k)}([s_1, s_2])\nonumber\\
	\geq&V_\alpha^{(k)}([s_1, s_2-z_2])+V_\alpha^{(k)}([s_1+z_1, s_2]).\nonumber
	\end{align}
	
	Notice that for any action $a_1, a_2, a_3, a_4$, we have $C([s_1+z_1, s_2-z_2], a_1)-C([s_1, s_2-z_2], a_2)=z_1$ and $C([s_1+z_1, s_2], a_3)-C([s_1, s_2],a_4)=z_1$. Define $\Delta$ be: \begin{align}
	\Delta\triangleq&\min_a\mathbb{E}_{s_1', s_2'}[V_\alpha^{(k)}([s_1', s_2'])|[s_1+z_1, s_2-z_2], a]\nonumber\\
	&+\min_a\mathbb{E}_{s_1', s_2'}[V_\alpha([s_1', s_2'])|[s_1, s_2], a]\nonumber\\
	&-\min_a\mathbb{E}_{s_1', s_2'}[V_\alpha^{(k)}([s_1', s_2'])|[s_1,s_2-z_2], a]\nonumber\\
	&-\min_a\mathbb{E}_{s_1', s_2'}[V_\alpha([s_1', s_2'])|[s_1+z_1, s_2], a].\nonumber
	\end{align} 
	To show the submodularity holds for $k+1$, it is suffices to prove $\Delta\geq 0$. Let $\pi^{(k+1)}(\mathbf{s})=\arg\min_a\mathbb{E}_\mathbf{s'}[V_\alpha^{(k)}(\mathbf{s}')|\mathbf{s}, a]$. For $s_1\neq0$ and $s_2-z_2\neq 0$, the proof is divided into two cases:
	
1). If $\pi^{(k+1)}([s_1+z_1, s_2-z_2])=\pi^{(k+1)}([s_1, s_2])=\tilde{a}$. With no loss of generality, assume $\tilde{a}=1$. Since $\tilde{a}$ may not be the optimum strategy for state $[s_1, s_2-z_2]$ and $[s_1+z_1, s_2]$, we have
\begin{align}
&\min_a\mathbb{E}_{s_1', s_2'}[V_\alpha^{(k)}([s_1', s_2'])|[s_1, s_2-z_2], a]\nonumber\\
\leq&\mathbb{E}_{s_1', s_2'}[V_\alpha^{(k)}([s_1', s_2'])|[s_1, s_2-z_2], \tilde{a}],\nonumber
\end{align}
and 
\begin{align}
&\min_a\mathbb{E}_{s_1', s_2'}[V_\alpha([s_1', s_2']|[s_1+z_1, s_2], a)]\nonumber\\
\leq&\mathbb{E}_{s_1', s_2'}[V_\alpha([s_1', s_2']|[s_1+z_1, s_2], \tilde{a})].\nonumber
\end{align} 
By plugging them into $\Delta$ we have:
		\begin{align}
		\Delta
		\geq&\mathbb{E}_{s_1', s_2'}[V_\alpha^{(k)}([s_1', s_2'])|[s_1+z_1, s_2-z_2], \tilde{a}]\nonumber\\
		&+\mathbb{E}_{s_1', s_2'}[V_\alpha^{(k)}([s_1', s_2'])|[s_1, s_2], \tilde{a}]\nonumber\\
		&-\mathbb{E}_{s_1', s_2'}[V_\alpha^{(k)}([s_1', s_2'])|[s_1,s_2-z_2], \tilde{a}]\nonumber\\
		&-\mathbb{E}_{s_1', s_2'}[V_\alpha^{(k)}([s_1', s_2'])|[s_1+z_1, s_2], \tilde{a}]\nonumber\\
		=&(1-p_1)(V_\alpha^{(k)}([s_1+z_1+1, s_2-z_2+1])\nonumber\\
		&+V_\alpha^{(k)}([s_1+1, s_2+1])-V_\alpha^{(k)}([s_1+1, s_2-z_2+1])\nonumber\\
		&-V_\alpha^{(k)}([s_1+z_1+1, s_2+1])).
		\end{align}
		Then according to the submodularity characteristic, we have $\Delta\geq 0$. The case when $\tilde{a}=2$ can be verified similarly.  

2). If $\pi^{(k+1)}([s_1+z_1, s_2-z_2])=a_1,\pi^{(k+1)}([s_1, s_2])=a_2, a_1\neq a_2$, with no loss of generality, suppose $a_1=1$ and $a_2=2$.
		
		If $p_1\leq p_2$, similar to the previous analysis, we have 
		\begin{align}
		&\min_a\mathbb{E}_{s_1', s_2'}[V_\alpha^{(k)}([s_1', s_2'])|[s_1, s_2-z_2], a]\nonumber\\
		\leq&\mathbb{E}_{s_1', s_2'}[V_\alpha^{(k)}([s_1', s_2'])|[s_1, s_2-z_2], a_1],\nonumber
		\end{align}
		and 
		\begin{align}
		&\min_a\mathbb{E}_{s_1', s_2'}[V_\alpha([s_1', s_2'])|[s_1+z_1, s_2], a]\nonumber\\
		\leq&\mathbb{E}_{s_1', s_2'}[V_\alpha([s_1', s_2'])|[s_1+z_1, s_2], a_2].\nonumber
		\end{align}
		 Then, $\Delta$ can be lower bounded by:
		\begin{align}
		\Delta
		\geq&\mathbb{E}_{s_1', s_2'}[V_\alpha^{(k)}([s_1', s_2'])|[s_1+z_1, s_2-z_2], a_1]\nonumber\\
		&+\mathbb{E}_{s_1', s_2'}[V_\alpha^{(k)}([s_1', s_2'])|[s_1, s_2], a_2]\nonumber\\&-\mathbb{E}_{s_1', s_2'}[V_\alpha^{(k)}([s_1', s_2'])|[s_1,s_2-z_2], a_1]\nonumber\\&-\mathbb{E}_{s_1', s_2'}[V_\alpha^{(k)}([s_1', s_2'])|[s_1+z_1, s_2], a_2]\nonumber\\
		=&(1-p_1)(V_\alpha^{(k)}([s_1+z_1+1, s_2-z_2+1])\nonumber\\
		&-V_\alpha^{(k)}([s_1+1, s_2-z_2+1]))\nonumber\\
		&+(1-p_2)(V_\alpha^{(k)}([s_1+1, s_2+1])\nonumber\\
		&-V_\alpha^{(k)}([s_1+z_1+1, s_2+1]))\nonumber\\
		=&(p_2-p_1)(V_\alpha^{(k)}([s_1+z_1+1, s_2-z_2+1])\nonumber\\
		&-V_\alpha^{(k)}([s_1+1, s_2-z_2+1]))\nonumber\\
		&+(1-p_2)(V_\alpha^{(k)}([s_1+z_1+1, s_2-z_2+1])\nonumber\\
		&+V_\alpha^{(k)}([s_1+1, s_2+1])\nonumber\\
		&-V_\alpha^{(k)}([s_1+1, s_2-z_2+1])\nonumber\\
		&-V_\alpha^{(k)}([s_1+z_1+1, s_2+1])).
		\end{align}	
		By monotonic, We have $V_\alpha^{(k)}([s_1+z_1+1, s_2-z_2+1])-V_\alpha^{(k)}([s_1+1, s_2-z_2+1])\geq 0$. Then combine the submodularity of $V_\alpha^{(k)}$, $\Delta\geq0$ can be verified.
		
		The case $p_1\geq p_2$ can be verified in the same way and is hence omitted.

		And for the case that $s_1=0$ or $s_2-z_2=0$, the proof needs some rectification similar to the proof in lemma 1, which is omitted here. Based on the above analysis, we have $\Delta\geq 0$ and the submodularity of $V_\alpha^{(k+1)}$ can be verified.
\end{section}
\begin{section}{Proof of Lemma 4}

Notice that $V_\alpha(s)$ is obtained by taking the minimum of all possible action sequence, hence, by choosing $a(t)=0$ all the time, we will formulate an upper bound on the $\alpha$-discounted problem. In this case, starting from any state $s$, according to the probability transfer function, the state of the decoupled bandit at time $t$ will satisfy $s(t)<s+t$. We can obtain the upper bound of $V_\alpha(s)$ by computing the total cost of applying this naive strategy:
\[V_\alpha(s)\leq\sum_{t=1}^\infty \alpha^{t-1}(s+t-1)=\frac{s}{1-\alpha}+\frac{1}{(1-\alpha)^2}.\]

Hence for every state $V_\alpha(s)<\infty$. Based on this characteristic, we can use a value iteration to approach the $\alpha$-discounted value function. Fixing $V_\alpha(0)=0$, the discounted value function obtained in the $(k+1)^{\text{th}}$ iteration can be obtained by:
\[V_\alpha^{(k+1)}(s)=\min_{a\in\{0,1\}}\{C(s,a)+\alpha\mathbb{E}_{s'}[V_\alpha^{(k)}(s')|s, a]\},\]
where $\mathbb{E}_{s'}[V_\alpha^{k}(s')|s, a]=\sum_{s'}V_\alpha^{(k)}(s')\text{Pr}(s'|s, a)$ denotes the expected $\alpha$-discounted function in the next time slot. We will then prove the monotonic characteristic of the value function by induction, suppose $V_\alpha^{(k)}(s)$ is a monotonically function of $s$, assume that $1\leq s_1<s_2$, then if $a=0$, according to the cost function $C(s, a)$ and the monotonic characteristic of $V_\alpha^{(k)}(s)$, we will have the following inequality:
\[C(s_1,0)+\alpha V_\alpha^{(k)}(s_1+1)\leq C(s_2, 0)+\alpha V_\alpha^{(k)}(s_2+1).\]

When $a=1$, starting from state $s_1\neq 0$, the bandit will evolve into state $0$, $1$ and state $s_1+1$ with probability $(1-\lambda)p$, $\lambda p$ and $1-p$, respectively; starting from state $s_2$, the bandit will evolve into state $0$, $1$ and state $s_2+1$ with probability $(1-\lambda)p$, $\lambda p$ and $1-p$. Then according to the monotonic characteristic of $V_\alpha^{(k)}(s)$, we will have
\[C(s_1, 1)+\alpha\mathbb{E}_{s'}[V_\alpha^{(k)}(s')|s_1, 1]\leq C(s_2, 1)+\alpha\mathbb{E}_{s'}[V_\alpha^{(k)}(s')|s_2, 1].\]

By taking the minimum over action set $\mathbb{A}$, the value of $V_\alpha^{(k+1)}(s)$ can be obtained and the following inequality holds:
\[V_\alpha^{(k+1)}(s_1)\leq V_\alpha^{(k+1)}(s_2)\]
Notice that when $k\rightarrow\infty$, we will have $V_\alpha^{(k)}(s)\rightarrow V_\alpha(s)$. Hence the monotonicity of the value function is proved.
\end{section}

\begin{section}{Derivations of Corollary 1}\label{Sec:C1}
		According to the transition probability of restless bandit, the state transition graph by applying threshold policy $\tau$ can be plotted as follows:
\begin{figure}[h]
	\centering
	\includegraphics[width=.45\textwidth]{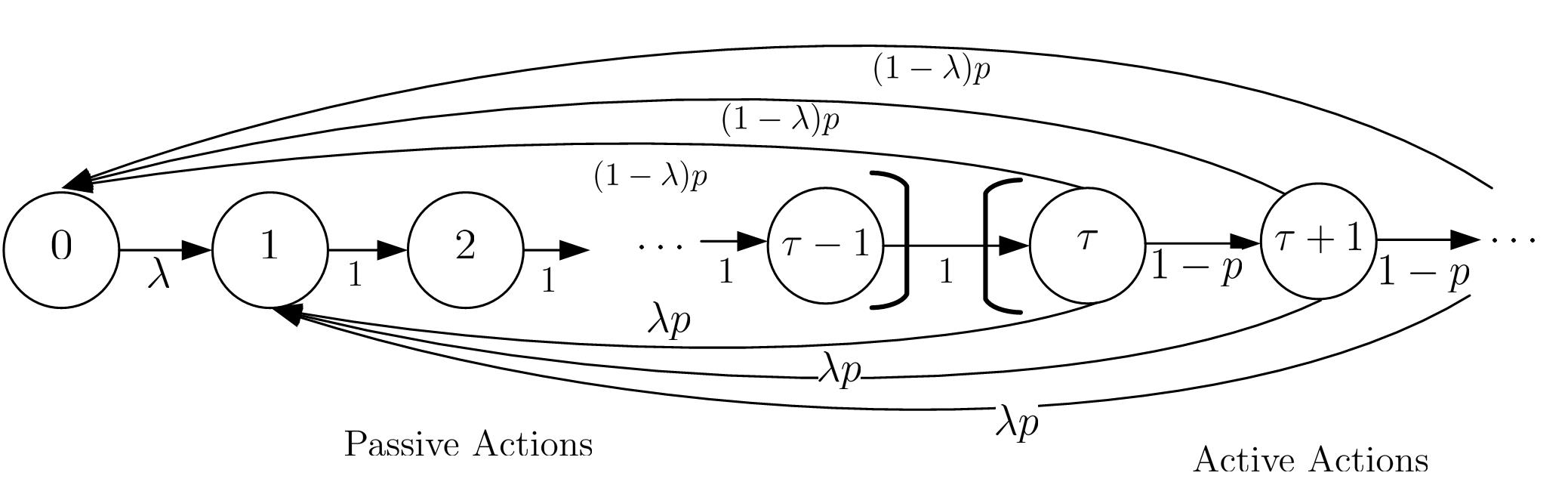}
	\caption{Probability transfer graph for threshold policy $\tau$, for states below $\tau$, the bandit remains passive; for states that are equal or larger than threshold $\tau$, the bandit becomes active. The transmission probability are denoted below the arrow.}
	\label{transfergraph}
\end{figure}

Denote $\xi_s^{(\tau)}$ to be the steady state distribution that the bandit is in state $s$ if a policy actives when $s\geq\tau$ but idles when $s<\tau$. Then according to the transition rule \eqref{banditprobability} and Fig.~\ref{transfergraph}, the relationship with the steady state distribution must satisfy:
		
		1) For $s<\tau$, the bandit remains passive, hence:
		\begin{subequations}
			\begin{equation}
			\xi_1^{(\tau)}=\cdots=\xi_\tau^{(\tau)}.
			\end{equation}

		2) For $s\geq \tau$, the bandit is chosen to be active. With probability $1-p$ the transmission fails and AoS grows to $s+1$:
			\begin{equation}
			\xi_{s+1}^{(\tau)}=(1-p)\xi_s^{(\tau)}, \forall s\geq \tau.\label{eq:xi1}
			\end{equation}
			
			The transmission succeeds with probability $p$, if a new update is sent to the BS with probability $\lambda$, then the bandit will go to state $s=1$:
			\begin{equation}
			\xi_1^{(\tau)}=\sum_{s=\tau}^\infty\lambda p \xi_s^{(\tau)}+\lambda \xi_0^{(\tau)}.
			\end{equation}
			
			If no update arrives and the transmission succeeds, the AoS at the next slot will go down to zero:
			\begin{equation}
			\xi_0^{(\tau)}=\sum_{s=\tau}^\infty(1-\lambda)p \xi_s^{(\tau)}+(1-\lambda)\xi_0^{(\tau)}.
			\end{equation}
			
			All these state distributions sum up to $1$, hence:
			\begin{equation}
				\sum_{s=0}^\infty\xi_s^{(\tau)}=1.\label{eq:xif}
			\end{equation}
		\end{subequations}
		
		Based on Eq.~\eqref{eq:xi1}-\eqref{eq:xif}, the steady state distribution $\xi_s^{(\tau)}$ can be computed as follows:
		\begin{equation}
		\xi_s^{(\tau)}=\begin{cases}
		\frac{1-\lambda}{\lambda}/\left(\frac{1-\lambda}{\lambda}+\tau+\frac{1}{p}-1\right),&s=0;\\
		1/\left(\frac{1-\lambda}{\lambda}+\tau+\frac{1}{p}-1\right),&1\leq s\leq \tau;\\
		\frac{(1-p)^{s-\tau}}{\left(\frac{1-\lambda}{\lambda}+\tau+\frac{1}{p}-1\right)},&s>\tau.
		\end{cases}
		\end{equation}
		Therefore, the average total cost $F(\tau, W)$ for given $W$ if threshold policy $\tau$ is employed can be computed:
		\begin{align}
		F(\tau,W)=&\sum_{s=0}^{\tau-1}s\xi_s^{(\tau)}+\sum_{s=\tau}^\infty(s+W)\xi_s^{(\tau)}\nonumber\\
		=&\sum_{s=1}^{\tau-1}s\xi_1^{(\tau)}+\sum_{s=\tau}^\infty(s+W)\xi_1^{(\tau)}(1-p)^{s-\tau}\nonumber\\
		=&\frac{\tau(\tau-1)}{2}\xi_1^{(\tau)}+\frac{\xi_1^{(\tau)}}{p}(\frac{1}{p}-1)+\frac{\xi_1^{(\tau)}}{p}(\tau+W).
		\end{align}
\end{section}

\begin{section}{Derivation of Corollary 2}\label{Sec:C2}
	According to our previous analysis, the optimum policy to solve the decoupled bandit possesses a threshold structure. Assume $\tau_\text{opt}$ is the optimum threshold, i.e., when AoS satisfies $0\leq s< \tau_\text{opt}$ the bandit idles and for $s\geq \tau_\text{opt}$ the bandit is activated. Denote $V(s)$ to be the differential cost-to-go function at state $s$, let $\beta$ be the optimum average cost. Then the Bellman equation can be written out as follows:
	\begin{align}
		V(s)+\beta=\min\{&(W+s)+(1-p)V(s+1)\nonumber\\
		&\hspace{0.5cm}+p(\lambda V(1)+(1-\lambda)V(0)), \nonumber\\
		&s+V(s+1)\}.		\label{eq:bandit} 
	\end{align}
	The above Bellman equation also implies, since it is optimum to activate the bandit in state $\tau_{\text{opt}}$, we have:
	\begin{subequations}
		\begin{equation}
			W+p((1-\lambda)V(0)+\lambda V(1))\leq pV(\tau_{\text{opt}}+1),\label{eq:activecon}
		\end{equation}
		and it is optimum to idle in state $\tau_{\text{opt}}-1$ implies:
		\begin{equation}
W+p((1-\lambda)V(0)+\lambda V(1))\leq pV(\tau_{\text{opt}}).\label{eq:idlecon}
\end{equation}		
	\end{subequations} The optimum threshold $\tau_{opt}$ is obtained by first writing out $V(\tau_{\text{opt}})$ and $\beta$ as a function of threshold $\tau_{\text{opt}}$. Then establish an equation to obtain $\tau_{\text{opt}}$. With no loss of generality, let us assume $\lambda V(1)+(1-\lambda)V(0)=0$. 
	
	To obtain $V(\tau_{\text{opt}})$, we first consider states satisfy $s\geq \tau_{opt}$. Since scheduling is the optimal action by assumption, according to the above Bellman equation Eq.~\eqref{eq:bandit}, the relationship between $V(s)$ and $V(s+1)$ is as follows:
	\begin{equation}
		V(s)=(-\beta+W+s)+(1-p)V(s+1). 
	\end{equation}
	
	Then, substitute $V(s+1)=(-\beta+W+s)+(1-p)V(s+2)$ into the above equation, we have:	\begin{align}
	V(s)=&(-\beta+W+s)+(1-p)(-\beta+W+s+1)\nonumber\\
	&+(1-p)^2V(s+2).
	\end{align} Repeating this procedure for $K$ times, we then have:
	\begin{equation}
	V(s)=\sum_{k=0}^{K-1}(1-p)^k(-\beta+W+s+k)+(1-p)^{K}V(s+K).
	\end{equation}
	
	Consider $K\rightarrow\infty$, since $\lim_{K\rightarrow\infty}(1-p)^KV(s+K)=0$, the differential cost-to-go function $V(\tau_{\text{opt}})$ can be obtained as follows:
	\begin{equation}
	V(s)=\frac{1}{p}(-\beta+W+s)+\frac{1-p}{p^2}.
	\label{eq:s>}		
	\end{equation}
	
	Recall Eq.~\eqref{eq:activecon},\eqref{eq:idlecon} and the assumption that $(1-\lambda)V(0)+\lambda V(1)=0$, since $\tau_{\text{opt}}$ is the optimum active threshold, we have:
	\begin{equation}
	V(\tau_{\text{opt}})\leq\frac{W}{p}<V(\tau_{\text{opt}}+1).
	\end{equation}
	
	Recall that $V(s)$ is monotonically increasing, the above inequality implies there exists a $\gamma\in[0, 1)$ such that $V(\tau_{\text{opt}}+\gamma)=\frac{W}{p}$. Plugging Eq.~\eqref{eq:s>} into $V(\tau_{\text{opt}}+\gamma)$, we can compute $\beta$ as follows:
	\begin{equation}
		\beta=\tau_{\text{opt}}+\gamma+\frac{1-p}{p}.
	\end{equation}
	Next, we consider states $s<\tau_{opt}$ and write $(1-\lambda)V(0)+\lambda V(1)$ as a function of $\tau_{\text{opt}}$. Since the optimum action is to idle when $s<\tau_{\text{opt}}$, according to the Bellman equation, we have:
	\begin{equation}
		V(s)=(-\beta+W+s)+V(s+1). 
	\end{equation}
	Substitute $V(s-1)=(-\beta+s-1)+V(s)$ into the above equation and repeat this procedure for $s-1$ times, we can then obtained the following equation about $V(1)$ and $V(\tau_{\text{opt}})$:
	\begin{equation}		V(1)=\frac{(\tau_{\text{opt}}-1)(-2\beta+\tau_{\text{opt}})}{2}+V(\tau_{\text{opt}}). 
		\label{eq:s<}
	\end{equation}
	The Bellman equation for $V(0)$ can be written out as follows:
	\[	V(0)=-\beta+(1-\lambda)V(0)+\lambda V(1),\]
	which implies
	\begin{equation}
	 V(0)=-\frac{\beta}{\lambda}+V(1). \label{eqopt1}
	\end{equation}
	
	Plugging the above equality into assumption $(1-\lambda)V(0)+\lambda V(1)=0$, the differential cost to go function $V(1)$ can be obtained:
	\begin{equation}
	V(1)=\frac{1-\lambda}{\lambda}\beta. \label{eq:threshold1}
	\end{equation}
	
	Plugging Eq.~\eqref{eqopt1} and Eq.~\eqref{eq:threshold1} into Eq.~\eqref{eq:s<}, we can write out $V(\tau_{\text{opt}})$ as follows:
	\begin{equation}
		V(\tau_{\text{opt}})=-\frac{(\tau_{\text{opt}}-1)(-2\beta+\tau_{\text{opt}})}{2}+\frac{1-\lambda}{\lambda}\beta. 
	\end{equation}
	
	Next, we establish a function about $\tau_{\text{opt}}$. Recall Eq.~\eqref{eq:s>}, when $s=\tau_{\text{opt}}$, we have another expression about $V(\tau_{\text{opt}})$:
	\begin{equation}
		V(\tau_{\text{opt}})=\frac{1}{p}(-\beta+W+\tau_{\text{opt}})+\frac{1-p}{p^2}.\label{eq:threshold2}
	\end{equation}
	
%
	
%
%
%
	Since the above two equalities should be equal, we can establish the following equation:
	\begin{align}
		-\frac{1-\lambda}{\lambda}\beta+\frac{(\tau_{\text{opt}}-1)(-2\beta+\tau_{\text{opt}})}{2}&\nonumber\\
		+\frac{1}{p}(-\beta+W+\tau_{\text{opt}})+\frac{1-p}{p^2}&=0. 
	\end{align}
	
	By substituting $\beta=\tau_{opt}+\gamma+\frac{1-p}{p}$ into the above equation, we have:
	\begin{align}
		-\frac{1-\lambda}{\lambda}(\tau_{\text{opt}}+\gamma+\frac{1-p}{p})&\nonumber\\
		+\frac{(\tau_{\text{opt}}-1)(-\tau_{\text{opt}}-2\gamma-2\frac{1-p}{p})}{2}+\frac{1}{p}(W-\gamma)&=0. 
	\end{align}
	
	The above equation is a quadratic equation about variable $\tau_{\text{opt}}$:
	\begin{align}
		&\frac{1}{2}\tau_{\text{opt}}^2+(\gamma+\frac{1}{p}+\frac{1}{\lambda}-\frac{5}{2})\tau_{\text{opt}}\\&-\frac{1}{p}(W-\gamma)
		+\frac{1-\lambda}{\lambda}(\gamma+\frac{1-p}{p})-2\frac{1-p}{p}-2\gamma=0.\nonumber
	\end{align}
	
	Since $\tau_{\text{opt}}$ is an integer and $\gamma\in[0, 1)$, we can finally obtain the threshold for fixed $W$, i.e., 
	\begin{align}
		\tau_{\text{opt}}&=\lfloor(\frac{5}{2}-\frac{1}{p}-\frac{1}{\lambda})+\nonumber\\
		&\sqrt{\left(\frac{5}{2}-\frac{1}{p}-\frac{1}{\lambda}\right)^2+2\left(\frac{W}{p}-\frac{1-\lambda}{\lambda}\frac{1-p}{p}\right)+2\frac{1-p}{p}}\rfloor.
	\end{align}
\end{section}

\end{document}